# Effective medium theory for photonic pseudospin-1/2 system

Neng Wang[1], Ruo-Yang Zhang[2], C. T. Chan[2] and Guo Ping Wang[1,*]

[1]*Institute of Microscale Optoelectronics, Shenzhen University, Shenzhen 518060, China*

[2]*Department of Physics, The Hong Kong University of Science and Technology, Clear Water Bay, Hong Kong, China*

Corresponding authors: *gpwang@szu.edu.cn

**Abstract:**

Photonic pseudospin-1/2 systems, which exhibit Dirac cone dispersion at Brillouin zone corners in analogy to graphene, have been extensively studied in recent years. However, it is known that a linear band crossing of two bands cannot emerge at the center of Brillouin zone in a two-dimensional photonic system respecting time reversal symmetry. Using a square lattice of elliptical magneto-optical cylinders, we constructed an unpaired Dirac point at the Brillouin zone center as the intersection of the second and third bands corresponding to the monopole and dipole excitations. Effective medium theory can be applied to the two linearly crossed bands with the effective constitutive parameters numerically calculated using the boundary effective medium approach. It is shown that only the effective permittivity approaches zero while the determinant of the nonzero effective permeability vanishes at the Dirac point frequency, showing a different behavior from the double-zero index metamaterials obtained from the pseudospin-1 triply degenerate points for time reversal symmetric systems. Exotic phenomena, such as the Klein tunneling and Zitterbewegung, in the pseudospin-1/2 system can be well understood from the effective medium description. When the Dirac point is lifted, the edge state dispersion near the Γ point can be accurately predicted by the effective constitutive parameters. We also further realized magneto-optical complex conjugate metamaterials for a wide frequency range by introducing a particular type of non-Hermittian perturbations which make the two linear bands coalescence to form exceptional points at the real frequency.

## I. Introduction

Pseudospin systems that exhibit linearly dispersive band structures have attracted much attention in the past decade due to their novel transport properties, such as Klein tunneling [1-3], Zitterbewegung [4-7], antilocalization [8, 9], and quantum Hall effect [10-13]. In classical waves, both pseudospin-1/2 Dirac cones and pseudospin-1 Dirac-like cones with an additional flat band crossing the triply degenerate point have been widely investigated. While the former is usually realized at the Brillouin zone corners in hexagonal lattices as photonic/phononic analogs of graphene [14-16], the latter can be achieved by the accidental triple degeneracy at the Brillouin zone center ($\Gamma$ point) [17-19]. In spite of their similar dispersions, the different pseudospin values of Dirac and Dirac-like points will result in fundamentally different transporting properties. For example, a pseudospin-1 system can support the super Klein tunneling effect [20-22], i. e., unity transmission for all incident angles, different from the Klein tunneling effect for Dirac cones where unity transmission is only guaranteed for the normal incidence. As such, pseudospin-1 system shows a different anomalous Anderson localization behavior from pseudospin-1/2 systems [23]. It has also been demonstrated that, in stark contrast to the Dirac cone which gives rise to a Berry phase of $\pi$ for loops encircling the Dirac point, the pseudospin-1 Dirac-like cone carries a zero Berry phase [24]. More importantly, because the Dirac-like point occurs at the Brillouin zone center, the effective medium theory can be rightfully applied [17, 18]. Effective medium theory provides a new perspective to understand the novel transporting properties of photonic pseudospin-1 systems [22]. Recently, it has also been reported that introducing non-Hermittian perturbations to the pseudospin-1 system could induce a ring of exceptional points (EPs) [25], and can be used for realizing the complex conjugate metamaterials (CCMs) [26].

In a time-reversal symmetry (TRS) preserved classical wave system, the bosonic TRS is incompatible with a 2D Dirac Hamiltonian located at time-reversal invariant momentum [27, 28]. Therefore, a single Dirac point cannot be formed at the Brillouin zone center if the TRS is unbroken. However, the local effective medium theory is only valid for waves with small Bloch wave vectors, making the effective medium theory inapplicable to describe the wave behavior at pseudospin 1/2 Dirac cones away from the Γ point. We note that although the double Dirac cone could be formed at the Γ point [29-31], they cannot be described as pseudospin-1/2 systems. The situation changes when the TRS of the system is broken [32], and a single Dirac point can be formed at the Γ point by simultaneously breaking the TRS and lattice symmetry [33, 34].

In this work, we constructed an unpaired Dirac point at the Γ point and at relatively low frequency using a square lattice of elliptical high-dielectric magneto-optical cylinders. The external magnetic field applied on the cylinders breaks TRS, and the elliptical shape breaks the $C_{4v}$ lattice symmetry. The Dirac point can be induced by tuning the magnitude of external magnetic field or the aspect ratio of the cylinder. Because the scatterings of the cylinders are dominated by the lowest order monopole and dipole excitations, we can apply the effective medium theory for the two linear bands and numerically calculate the effective constitutive parameters using the boundary effective medium approach. We found that although the effective permittivity approaches zero at the Dirac point frequency, the effective permeability remains as a nonzero tensor with gyromagnetic response, which is different from traditional double-zero index metamaterials corresponding to the pseudospin-1 systems whose permittivity and permeability tend to zero simultaneously at the Dirac-like point frequency [17]. However, it is interesting that the determinant of the effective permeability is zero at the Dirac point frequency. Therefore, this metamaterial can be regarded as a generalized form of double-zero index metamaterial with the effective permittivity as well as the determinant of the permeablity tensor to be zero.

Some exotic phenomena, such as the Klein tunneling and Zitterbewegung effects, were observed in the homogenized effective medium. We found that the Klein tunneling effect can be understood from the mechanism of impedance match, and the Zitterbewegung effect can be explained from the interference of the electromagnetic (EM) waves with different polarizations on the two linearly crossing bands. Also, we investigated the edge states within the gap of the lifted Dirac cones at the interface of two magneto-optic photonic crystals with equal but opposite magnetic fields. We show that the dispersion of the surface modes closing to the Γ point produced by the two adjoined effective magneto-optic media is consistent with that of the real photonic crystal structure. Therefore, the effective constitutive parameters can reflect not only the bulk but also the surface transporting properties of this pseudospin-1/2 photonic crystal.

In addition, by introducing a particular type of non-Hermittian perturbations to the photonic crystal, we achieved a ring of EPs encircling the original Dirac points and demonstrated that such structure forms a magneto-optical CCMs valid for a wide frequency range. In contrast to conventional CCMs [26, 35-37], the effective permittivity of a magneto-optical CCM is not complex conjugate to any effective permeability tensor components. Nevertheless, the magneto-optical CCMs remain one of the core properties of conventional CCMs, i.e. their effective refractive indices are purely real. Similar to conventional CCMs, the magneto-optical CCMs can also support both coherent perfect absorbing (CPA) and lasing modes.

## II. Unpaired Dirac cone at Γ point by accidental degeneracy

The system we considered is a photonic crystal composed of identical elliptical magneto-optical cylinders arranged into a square lattice embedded in air, as shown in the inset of Fig. 1(a). The relative permittivity of the cylinder is $\varepsilon_r = 12.5$. When the external magnetic field is applied along the $z$ direction, the in-plane relative permeability tensor of the cylinder is given by

$$\vec{\mu}_c = \begin{pmatrix} \mu_r & -i\mu_k \\ i\mu_k & \mu_r \end{pmatrix}, (1)$$

where $\mu_r = 1$ is used throughout the paper for simplicity and $\mu_k$ depends on the magnitude of the external magnetic field. The semi-minor and semi-major axis of the cylinder are respectively, $r_a = 0.17a, r_b = 1.205 r_a$, where $a$ is the lattice constant. When there is no external magnetic field, namely $\mu_k = 0$, the band diagram for $E_z$ polarized EM field is calculated and shown in Fig. 1(a). We can see that there is small band gap between the second and third bands near the Γ point. A fine tuning of the aspect ratio $r_b / r_a$ can lead to an accidental degeneracy of the second and third bands, forming a semi-Dirac point at Γ point where the two bands intersect linearly along Γ-Y direction but touch quadratically along Γ-X direction [38]. Another way to make the second and third bands degenerate is by applying external static magnetic field. As we can see from Fig. 1(b), for $\mu_k = 0.3$, the second and third bands become degenerate at the Γ point. However, the degenerate point now is not a semi-Dirac point but a Dirac point, as the linear dispersion in all directions are clearly seen from Fig. 1 (c). The frequency for Dirac point is at $\omega_D a / (2\pi c) = 0.55$.

The external static magnetic field breaks the time-reversal symmetry (TRS) of the photonic crystal. To investigate how the Dirac point is formed, we now formulate the Hamiltonian of the TRS broken ($\mu_k \neq 0$) system using the eigenstates of the TRS preserved ($\mu_k = 0$) system in conjunction with the wave equation for the magnetic field $\mathbf{H} = (H_x, H_y)$,

$$\nabla_\perp \times (\frac{1}{\varepsilon(\mathbf{r})} \nabla_\perp \times \mathbf{H}) = -\left(\frac{\omega}{c}\right)^2 \vec{\mu}(\mathbf{r}) \cdot \mathbf{H}, (2)$$

where $\nabla_\perp = (\partial_x, \partial_y)$ is the in-plane nabla operator, $c$ is the light speed in vacuum, and $\varepsilon(\mathbf{r}), \tilde{\mu}(\mathbf{r})$ are the position dependent relative permittivity and permeability tensors. According to the Bloch theorem, the magnetic field of the TRS preserved system can be written as $\mathbf{H} = e^{i\mathbf{k}\cdot\mathbf{r}} \mathbf{u}_{n\mathbf{k}}(\mathbf{r})$, where $\mathbf{k}$ is the Bloch wave vector, $\mathbf{u}_{n\mathbf{k}}(\mathbf{r})$ is the normalized cell function of the $n$th band. $\{\mathbf{u}_{n\mathbf{k}}(\mathbf{r})\}$ forms a complete orthogonal set due to the Hemiticity of the system. Therefore the eigenstate of the TRS broken system $\tilde{\mathbf{u}}_{n\mathbf{k}}(\mathbf{r})$ can be expanded in series of $\{\mathbf{u}_{n\mathbf{k}}(\mathbf{r})\}$ as

$$\tilde{\mathbf{u}}_{n\mathbf{k}}(\mathbf{r}) = \sum_n \alpha_{nm} \mathbf{u}_{m\mathbf{k}}(\mathbf{r}), \quad (3)$$

where $\alpha_{nm}$ is the expansion coefficient to be determined. Substituting Eq. (3) into Eq. (2), we have

$$-\sum_m \mu_r(\mathbf{r}) \alpha_{nm} \frac{\omega_m^2}{c^2} \mathbf{u}_{m\mathbf{k}} + \left(\frac{\omega}{c}\right)^2 \begin{pmatrix} \mu_r(\mathbf{r}) & -i\mu_k(\mathbf{r}) \\ i\mu_k(\mathbf{r}) & \mu_r(\mathbf{r}) \end{pmatrix} \cdot \sum_m \alpha_{nm} \mathbf{u}_{m\mathbf{k}} = 0, \quad (4)$$

where $\omega_m$ is the frequency of the $m$th band of the TRS preserved system (shown in Fig.1(a)). Inner product $(\mathbf{u}_{n\mathbf{k}}^*)^T$ from left and integrate within a unit cell, using the orthogonality of the eigenstates $\{\mathbf{u}_{n\mathbf{k}}(\mathbf{r})\}$, we arrive at

$$-\sum_m \{\delta_{nm} \frac{\omega_m^2}{c^2} \iint \mu_r |\mathbf{u}_{m\mathbf{k}}|^2 \, d\mathbf{r} + \frac{\omega^2}{c^2} \alpha_{nm}[-i\mu_k \underbrace{\iint_{S_c} \mathbf{u}_{n\mathbf{k}}^* \times \mathbf{u}_{m\mathbf{k}} \cdot \hat{z} d\mathbf{r}}_{P_{nm}} + \delta_{nm} \iint \mu_r |\mathbf{u}_{m\mathbf{k}}|^2 \, d\mathbf{r}\} = 0, \quad (5)$$

where $\delta_{nm}$ is the Kronecker delta function, $S_c$ denotes the region inside the cylinder. By normalization, the first and third integrals in Eq. (5) are equal to 1. Eq. (5) then can be formulated into an eigenvalue problem as

$$\hat{H}_1^{-1} \cdot \hat{H}_2 \cdot \vec{\psi} = \left(\frac{\omega}{c}\right)^2 \vec{\psi}, \quad (6)$$

where $\vec{\psi} = \{\alpha_{nm}\}^T$,

$$H_{1,nm} = \delta_{nm} - i\mu_k p_{nm}, \quad H_{2,nm} = \delta_{nm} \frac{\omega_m^2}{c^2}. \quad (7)$$

Therefore the Hamiltonian of the TRS broken system is $\hat{H}_1^{-1} \cdot \hat{H}_2$ and the coupling strengths between different bands are proportional to $\mu_k$. Considering only the second, third and fourth bands in Eq. (6), the eigenvalues $(\omega/c)^2$ at $\mathbf{k}=0$ as function of $\mu_k$ is shown in Fig. 1 (d). We can see that while the second band is almost invariant, the third and fourth bands repel each other, which induces the third band to move downward as $\mu_k$ increases. The movement of the third band leads to a band inversion between the second and third bands occurring at about $\mu_k = \pm 0.3$. As a result, the Dirac point is formed mainly due to the repulsion behavior between the third and fourth bands caused by the magneto-optical perturbation $\mu_k$. According to Eq. (6), the values of $\mu_k$ causing the emergence of the Dirac point for different aspect ratios $r_b/r_a$ can be obtained.

### III. Effective constitutive parameters obtained using the boundary effective medium approach

#### A. The boundary effective medium approach for magneto-optical metamaterials

As the Dirac point is located at the Brillouin zone center, and the scattering properties of the cylinder near $\omega_D$ are dominated by the monopole and dipole resonant modes, we can apply the effective medium theory to the two linearly crossing bands nearby the Brillouin zone center.

To obtain the effective constitutive parameters, we introduce the boundary effective medium approach [38]. The main idea of boundary effective medium approach is using the boundary averaged fields of the eigenstates to describe the effective fields, which is different from traditional homogenization method that uses the volume averaged fields [39]. The latter is good for the lowest band but usually performs less satisfactorily for the higher bands because of the strong spatial dispersion [40].

The boundary effective medium approach was originally proposed for calculating the effective parameters of nonmagnetic metamaterials. Here we extend the method for magneto-optical metamaterials. The effective constitutive parameters are calculated according to the constitutive relations,

$$\bar{D}_z = \varepsilon_{ez}\varepsilon_0 \bar{E}_z, \quad \bar{\mathbf{B}} = \mu_0 \vec{\mu}_e \cdot \bar{\mathbf{H}} = \mu_0 \begin{pmatrix} \mu_{ex} & -i\mu_{ek} \\ i\mu_{ek} & \mu_{ey} \end{pmatrix} \cdot \bar{\mathbf{H}}, \quad (8)$$

where $\bar{E}_z, \bar{D}_z, \bar{\mathbf{H}}$ and $\bar{\mathbf{B}}$ are the averaged fields, $\varepsilon_e, \vec{\mu}_e$ are the effective permittivity and effective permeability tensor. For the eigenstate with Bloch vectors along ΓX direction, the averaged fields are calculated as

$$\bar{E}_z(k_x) = \frac{1}{2a}\int_0^a E_z(x=0)dy + \frac{1}{2a}\int_0^a E_z(x=a)dy,$$

$$\bar{H}_x(k_x) = \frac{1}{2a}\int_0^a H_x(y=0)dx + \frac{1}{2a}\int_0^a H_x(y=a)dx, \quad (9)$$

$$\bar{H}_y(k_x) = \frac{1}{2a}\int_0^a H_y(x=0)dy + \frac{1}{2a}\int_0^a H_y(x=a)dy,$$

where $x=0,a$ or $y=0,a$ denote the boundaries of the unit cell. The averaged electric displacement and the magnetic induced fields are then obtained according to the Maxwell's equations in conjunction with the finite difference method (valid as $ka \ll 1$)

$$\overline{D}_z(k_x) = -\frac{1}{i\omega a^2}\int_0^a [H_y(x=a) - H_y(x=0)]dy,$$

$$\overline{B}_x(k_x) = \frac{1}{i\omega a^2}\int_0^a [E_z(y=a) - E_z(y=0)]dx, \quad (10)$$

$$\overline{B}_y(k_x) = -\frac{1}{i\omega a^2}\int_0^a [E_z(x=a) - E_z(x=0)]dy.$$

Similarly, for the eigenstate with Bloch vectors along ΓY direction, the averaged fields are calculated as

$$\overline{E}_z(k_y) = \frac{1}{2a}\int_0^a [E_z(y=0) + E_z(y=a)]dx,$$

$$\overline{H}_x(k_y) = \frac{1}{2a}\int_0^a [H_x(y=0) + H_x(y=a)]dx, \quad (11)$$

$$\overline{H}_y(k_y) = \frac{1}{2a}\int_0^a [H_y(x=0) + H_y(x=a)]dy,$$

and the averaged electric displacement and the magnetic induced fields are

$$\overline{D}_z(k_y) = \frac{1}{i\omega a^2}\int_0^a [H_x(y=a) - H_x(y=0)]dx,$$

$$\overline{B}_x(k_y) = \frac{1}{i\omega a^2}\int_0^a [E_z(y=a) - E_z(y=0)]dx, \quad (12)$$

$$\overline{B}_y(k_y) = \frac{1}{i\omega a^2}\int_0^a [E_z(x=0) - E_z(x=a)]dy.$$

According to the Bloch theorem, we know that $\overline{B}_x(k_x) = \overline{B}_y(k_y) = 0$. Substituting Eqs. (9)-(12) into Eq. (8), we obtain that

$$\varepsilon_0 \varepsilon_{ez} = \frac{\overline{D}_z(k_x)}{\overline{E}_z(k_x)}, \qquad \varepsilon_0 \varepsilon_{ez} = \frac{\overline{D}_z(k_y)}{\overline{E}_z(k_y)},$$

$$\mu_0 \mu_{ex} = \frac{\overline{B}_x(k_y)\overline{H}_y(k_x)}{\overline{H}_x(k_y)\overline{H}_y(k_x) - \overline{H}_x(k_x)\overline{H}_y(k_y)},$$

$$\mu_0 \mu_{ek} = -i \frac{\overline{B}_x(k_y)\overline{H}_x(k_x)}{\overline{H}_x(k_y)\overline{H}_y(k_x) - \overline{H}_x(k_x)\overline{H}_y(k_y)}, (13)$$

$$\mu_0 \mu_{ey} = \frac{\overline{B}_y(k_x)\overline{H}_x(k_y)}{\overline{H}_x(k_y)\overline{H}_y(k_x) - \overline{H}_x(k_x)\overline{H}_y(k_y)},$$

$$\mu_0 \mu_{ek}' = i \frac{\overline{B}_y(k_x)\overline{H}_y(k_y)}{\overline{H}_x(k_y)\overline{H}_y(k_x) - \overline{H}_x(k_x)\overline{H}_y(k_y)}.$$

Although $\varepsilon_{ez}$ is overdetermined by the first two equations, they actually produce the same numerical results, as we will see in the following. Also, for Hermitian systems, $\mu_{ek}$ and $\mu_{ek}'$ must be equal.

The effective constitutive parameters calculated using Eq. (13) for the two linear bands are shown in Fig. 2. We can see that the effective permittivity $\varepsilon_{ez}$ obtained from the first two formulas in Eq. (13) are indeed the same. Also, the equality of $\mu_{ek}$ and $\mu_{ek}'$ is necessary as the system is Hermitian. It is worth noticing that at the Dirac point frequency $\omega_D$, only the effective permittivity goes to zero, while the components of the effective permeability tensor are all negative. This result is markly different from the situation of a pseudospin-1 Dirac-like point where both the effective permittivity and permeability approach zero at the accidental triply degenerate frequency [17]. We also see that, in the vicinity of the Dirac point, all the effective constitutive parameters change linearly with respect to $\omega - \omega_D$, and can be written as

$$\varepsilon_{ez} = \gamma_z \Delta\tilde{\omega}, \mu_{ex} = \mu_{x0} + \gamma_x \Delta\tilde{\omega}, \mu_{ey} = \mu_{y0} + \gamma_y \Delta\tilde{\omega}, \mu_{ek} = \mu_{k0} + \gamma_k \Delta\tilde{\omega}, (14)$$

where $\Delta\tilde{\omega} = (\omega - \omega_D)a/c$ is the dimensionless frequency displacement from $\omega_D$, and the parameters $\gamma_x, \gamma_y, \gamma_z, \gamma_k, \mu_{x0}, \mu_{y0}, \mu_{k0}$ are obtained by linear fitting the numerical data in Fig.2 and their values are given in the figure caption.

For the homogenized medium, the refractive index is calculated by $kc/\omega$, where $k$ is the wavenumber and $\omega$ is the circular frequency. According to the Maxwell's equation,

$$i\mathbf{k}\times\overline{\mathbf{E}} = i\omega\mu_0\bar{\bar{\mu}}_e \cdot \overline{\mathbf{H}}, \quad i\mathbf{k}\times\overline{\mathbf{H}} = -i\omega\varepsilon_0\varepsilon\overline{\mathbf{E}}, (15)$$

we can obtain the following secular equation as

$$\begin{pmatrix} -i\dfrac{\omega}{c}\mu_{ex} & -\dfrac{\omega}{c}\mu_{ek} & ik_y \\ \dfrac{\omega}{c}\mu_{ek} & -i\dfrac{\omega}{c}\mu_{ey} & -ik_x \\ -ik_y & ik_x & i\dfrac{\omega}{c}\varepsilon_{ez} \end{pmatrix} \begin{pmatrix} Z_0\overline{H}_x \\ Z_0\overline{H}_y \\ \overline{E}_z \end{pmatrix} = 0, (16)$$

where $Z_0 = \sqrt{\mu_0/\varepsilon_0}$ is the impedance of vacuum. The iso-frequency surface contour is obtained from the condition

$$\det\begin{pmatrix} -i\dfrac{\omega}{c}\mu_{ex} & -\dfrac{\omega}{c}\mu_{ek} & ik_y \\ \dfrac{\omega}{c}\mu_{ek} & -i\dfrac{\omega}{c}\mu_{ey} & -ik_x \\ -ik_y & ik_x & i\dfrac{\omega}{c}\varepsilon_{ez} \end{pmatrix} = 0. (17)$$

which corresponds to the nontrivial solutions of Eq. (16).

By, respectively, setting $k_y = 0$ and $k_x = 0$, and solving Eq. (17) we obtained the refractive indices for waves propagating along $x$ and $y$ directions as

$$n_x = \sqrt{\varepsilon_{ez}\frac{\mu_{ex}\mu_{ey}-\mu_{ek}^2}{\mu_{ex}}}, \quad n_y = \sqrt{\varepsilon_{ez}\frac{\mu_{ex}\mu_{ey}-\mu_{ek}^2}{\mu_{ey}}}. (18)$$

And the corresponding impedances are given by

$$Z_{ex} = \frac{n_x}{\varepsilon_{ez}} = \sqrt{\frac{\mu_{ex}\mu_{ey} - \mu_{ek}^2}{\varepsilon_{ez}\mu_{ex}}}, \quad Z_{ey} = \frac{n_y}{\varepsilon_{ez}} = \sqrt{\frac{\mu_{ex}\mu_{ey} - \mu_{ek}^2}{\varepsilon_{ez}\mu_{ey}}}. \quad (19)$$

**B. Hamiltonian based on Maxwell's equations**

Substituting Eq. (14) into Eq. (15), we obtain

$$\tilde{k}_y \bar{E}_z = (\tilde{\omega}_D + \Delta\tilde{\omega})(\mu_{x0} + \gamma_x \Delta\tilde{\omega})Z_0 \bar{H}_x + i(\tilde{\omega}_D + \Delta\tilde{\omega})(\mu_{k0} + \gamma_k \Delta\tilde{\omega})Z_0 \bar{H}_y,$$
$$-\tilde{k}_x \bar{E}_z = (\tilde{\omega}_D + \Delta\tilde{\omega})(\mu_{y0} + \gamma_y \Delta\tilde{\omega})Z_0 \bar{H}_y - i(\tilde{\omega}_D + \Delta\tilde{\omega})(\mu_{k0} + \gamma_k \Delta\tilde{\omega})Z_0 \bar{H}_x, \quad (20)$$
$$\tilde{k}_x Z_0 \bar{H}_y - \tilde{k}_y Z_0 \bar{H}_x = -(\tilde{\omega}_D + \Delta\tilde{\omega})\gamma_z \Delta\tilde{\omega}\bar{E}_z$$

where $\tilde{\mathbf{k}} = \mathbf{k}a$ is the dimensionless wave vector. Ignoring the higher order terms $(\Delta\tilde{\omega})^2$, we obtain the eigenvalue problem as

$$\begin{pmatrix} \mu_{x0} + \gamma_x \tilde{\omega}_D & i(\mu_{k0} + \gamma_k \tilde{\omega}_D) & 0 \\ -i(\mu_{k0} + \gamma_k \tilde{\omega}_D) & \mu_{y0} + \gamma_y \tilde{\omega}_D & 0 \\ 0 & 0 & \gamma_z \tilde{\omega}_D \end{pmatrix}^{-1} \begin{pmatrix} -\tilde{\omega}_D \mu_{x0} & -i\tilde{\omega}_D \mu_{k0} & \tilde{k}_y \\ i\tilde{\omega}_D \mu_{k0} & -\tilde{\omega}_D \mu_{y0} & -\tilde{k}_x \\ \tilde{k}_y & -\tilde{k}_x & 0 \end{pmatrix} \begin{pmatrix} Z_0 \bar{H}_x \\ Z_0 \bar{H}_y \\ \bar{E}_z \end{pmatrix} = \Delta\tilde{\omega} \begin{pmatrix} Z_0 \bar{H}_x \\ Z_0 \bar{H}_y \\ \bar{E}_z \end{pmatrix}.$$

(21)

From Eq. (21), it is easy to find that there is an unpaired Dirac point at $\mathbf{k} = 0$ and $\Delta\tilde{\omega} = 0$ only when the determinant of the effective permeability is reduced to zero, namely $\mu_{k0}^2 = \mu_{x0}\mu_{y0}$. In Fig. 3(a), the band structure of the effective medium calculated by Eq. (21) are compared with the band structure of the photonic crystal calculated using the finite element method software COMSOL [41]. We see that the effective parameters can accurately reproduce the conical band dispersion. However, since the effective medium theory does not take into account the fourth band of the photonic crystal, Eq. (21) cannot produce the fourth band correctly. The Dirac point will be lifted if $\mu_{k0}^2 \neq \mu_{x0}\mu_{y0}$, which can be verified from the numerical results shown in Figs. 3(b) and 3(c). For the special case where $\gamma_x = \gamma_y, \mu_{x0} = \mu_{y0}$, we can formulate the Hamiltonian for the two crossing bands in terms of the Pauli matrices using the k.p method, and see clearly that the band gap size is proportional to the detuning $\delta = |\mu_{k0}| - \sqrt{\mu_{x0}\mu_{y0}}$, see the Appendix A.

## IV. Exotic phenomena in pseudospin-1/2 system

### A. Klein tunneling

It is well-known that artificial systems with nodal points, such as Dirac and Weyl nodes, can serve as ideal platforms to investigate various novel physical phenomena that were unique properties of real spinful particles. One such exotic phenomena is Klein tunneling, which refers to the total transmission effect of spinful (quasi-)particles penetrating a wide potential barrier. Klein tunneling was first envisioned in relativistic quantum physics [42] and experimentally observed in graphene which exhibits pseudospin-1/2 carrier dynamics [1-3]. Photonic analog of this 2D material, i.e., the photonic graphene, provides a platform for observing the Klein tunneling in the classical wave system [43, 44]. It has also been demonstrated that the "super" Klein tunneling effect, i.e., total transmission for all incident angles, is supported in the photonic pseudospin-1 systems which exhibit Dirac-like cone band dispersions [22], the mechanism of which has also been understood from the concept of "complementary materials" in EM wave theory [22]. In what follows, we show that Klein tunneling can also be realized in our unpaired Dirac cone system, and provide a new understanding from the impedance of magneto-optical zero-index medium.

The configuration for observing the Klein tunneling effect in effective medium is shown in Fig. 4, where a plane wave is incident on an effective medium slab from the effective medium background. We assume that the effective constitutive parameters of both the background medium and slab are given by Eq. (14), whereas the Dirac point frequencies of the slab and background are slightly different, which are $\omega_D$ and $\omega_D' = 0.95\omega_D$, respectively. The difference between the Dirac point frequencies of the slab and background induces a square potential barrier $V = \omega_D - \omega_D' = 0.05\omega_D$. In Fig. 5, we showed the transmittance of the normally incident EM plane waves varying with frequencies. Comparing with the band dispersions of the effective medium slab (black lines) and background (orange lines) plotted in the left panel, we can see that the transmittance is close to unity within the linear band dispersion region, and drops dramatically at higher frequencies where the higher-order correction to band dispersion contributes.

The Klein tunneling effect can be understood from the scattering theory of EM wave. According to the definition Eq. (19), the impedance of the effective medium is given by

$$Z_{ex}^2 = \frac{\mu_{ex}\mu_{ey} - \mu_{ek}^2}{\varepsilon_{ez}\mu_{ex}} = \frac{\mu_{x0}\mu_{y0} - \mu_{k0}^2}{\gamma_e \mu_{x0} \Delta\tilde{\omega}} + \frac{\mu_{x0}^2 \gamma_y + \mu_{k0}^2 \gamma_x - 2\gamma_k \mu_{x0}\mu_{k0}}{\gamma_z \mu_{x0}^2} + o(\Delta\tilde{\omega}),$$
$$Z_{ey}^2 = \frac{\mu_{ex}\mu_{ey} - \mu_{ek}^2}{\varepsilon_{ez}\mu_{ex}} = \frac{\mu_{x0}\mu_{y0} - \mu_{k0}^2}{\gamma_e \mu_{y0} \Delta\tilde{\omega}} + \frac{\mu_{y0}^2 \gamma_x + \mu_{k0}^2 \gamma_y - 2\gamma_k \mu_{y0}\mu_{k0}}{\gamma_z \mu_{y0}^2} + o(\Delta\tilde{\omega}). \tag{22}$$

We can see that when $\mu_{k0}^2 = \mu_{x0}\mu_{y0}$ the impedance of the effective medium is independent of frequency in the limit of vanishing $\Delta\tilde{\omega}$. Therefore, provided that when the two Dirac point frequencies $\omega_D$ and $\omega_D'$ are close enough, the impedance match guarantees the unity transmission.

When the frequency of the incident wave is $\omega = (\omega_D + \omega_D')/2$, super Klein tunneling will occur in the pseudospin-1 photonic system [22]. In contrast to the effective double-zero-index medium associated with an isotropic Dirac-like cone, the effective magneto-optical zero index medium we investigate here does not give unity transmittance when the beam is obliquely incident, as shown by the red circles shown in Fig. 6. The super Klein tunneling effect is caused by the fact that the barrier slab forms the complementary material of the background, i.e., the permittivity and permeability of the barrier slab take equal but opposite values to those of the background. However, the effective medium with effective constitutive parameters in Eq. (14) does not fulfil the condition for complementary material due to the nonzero $\mu_{x0}, \mu_{y0}$ and $\mu_{k0}$.

**B. Zitterbewegung**

Zitterbewegung, which represents the trembling motion of quasiparticles or wavepackets described by the Dirac equation, is another exotic phenomenon well known in pseudospin-1/2 systems [4-7]. To investigate the Zitterbewegung effect, we consider the propagation of a wavepacket in our pseudospin-1/2 effective medium. We suppose the wavepacket is a plane wave in $x$ direction with a certain $k_x$ but has a Gaussian type profile with a finite beam width along the $y$ direction. Then the expression of EM fields in the real space at time $t$ is given by

$$\begin{pmatrix} Z_0 \bar{H}_x \\ Z_0 \bar{H}_y \\ \bar{E}_z \end{pmatrix} = e^{ik_x x} \int_{-\infty}^{\infty} dk_y \exp[-\left(\frac{k_y}{\delta_y}\right)^2 + ik_y y] \sum_{i=1,2} f_i(k_y) \vec{\psi}_i(k_y) e^{-i\omega_i t}, \quad (23)$$

where $\delta_y$ is the width of wavepaket in the k space, $\omega_i$ are the $k_y$-dependent frequencies for the lower ($i$=1) and upper ($i$=2) bands of the Dirac cone, $\vec{\psi}_i(k_y)$ is the corresponding normalized eigen-polarization which can be determined according to Eq. (21). According to the mechanism of Zitterbewegung, the oscillation trajectory of wavepacket stems from the interference between the states on the two bands of the Dirac cone, and hence we should require both the two eigenstates on the two linear bands, $\vec{\psi}_1(k_y)$ and $\vec{\psi}_2(k_y)$, to appear simultaneously in the superposed field, namely both $f_1(k_y)$ and $f_2(k_y)$ should be nonzero, in order to achieve a nontrivial result. In the following, we would assume equally weighted superposition of the two bands, i.e., $f_1(k_y) = f_2(k_y) = 1$, which can maximize the amplitude of the Zitterbewegung. For a quasi-monochromatic wavepacket, the effective energy density of the EM wave averaged in a quasi-time-period for different location $y$ and time $t$ can be calculated by [45]

$$w_{eff}(y,t) \sim \frac{1}{2}\bar{E}_z^*(y,t) \cdot \left.\frac{d(\omega\varepsilon_{ez})}{d\omega}\right|_{\omega=\omega_D} \cdot \bar{E}_z(y,t) + \frac{1}{2}\bar{\mathbf{H}}_t^*(y,t) \cdot \left.\frac{d(\omega\vec{\mu}_e)}{d\omega}\right|_{\omega=\omega_D} \cdot (\bar{\mathbf{H}}_t(y,t))^T, \quad (24)$$

where $\bar{\mathbf{H}}_t = Z_0(\bar{H}_x, \bar{H}_y)$.

In Fig. 7(a), we plotted the effective energy density distribution for the wavepacket with $k_x = 0.3/a$ and $\delta_y = 0.1/a$. The trembling behavior is clearly seen, especially when the diffraction effect is weak for $t < 40\omega_D/\pi$. The centroid of a wavepacket can be defined according to the effective energy density as $y_c = \int w_{eff}(y,t) y dy / \int w_{eff}(y,t) dy$. Fig. 7(b) shows clearly the trembling motion of the wavepacket centroid $y_c$ with time $t$, where the decay of the oscillation amplitude arises from the diffraction of the wave packet in time.

## C. Surface states

The gaping of the Dirac points due to Hermittian perturbations can induce the quantum Hall effect [12, 13], quantum spin Hall effect [46], quantum valley Hall effect [47, 48] and even the hybridization of these effects [32, 49]. A consequence of these effects is the emergency of topological nontrivial surface states which possess novel transporting and scattering properties, such as one-way transport and backscattering immunity against local disorders [12, 13, 32, 46-49]. The number of topologically nontrivial surface states are determined by the topological invariants of the isolated bands which can be calculated using the bulk eigenstates [12], such as the Chern number, spin-Chern number and valley Chern number. However, because the bulk-edge correspondance only gives the net number of one-ay modes, concrete dispersion relations of the surface states cannot be determined by topological principles if more details are not given.

The effective medium theory reflects the macroscopic properties of the metamaterial. When using effective medium theory, some questions naturally arise, such as (1) whether the effective constitutive parameters can also be used to calculate the surface state dispersion even when the waves are not penetrating so deep inside the metamaterial, and (2) whether the effective constitutive parameters can give more information than the topological numbers.

To answer these questions, we consider the formations of surface states at the interface of two metamaterials composed of magneto-optical cylinders magnetized in an opposite manner. The surface states of the real structure are calculated using the supercell approach, as shown in Fig. 8. The supercell contains 30 layers along *x* direction and one layer along *y* direction, and periodic boundary conditions are imposed in both directions. Because of the magnetization setting, the upper half and lower half cylinders have opposite magneto-optical couplings $\mu_k$, forming the inner and outer interfaces which correspond to the central line and the top and bottom boundaries of the supercell, see Fig. 8. When $|\mu_k| \neq 0.3$, the Dirac point is lifted and a band gap forms. For $|\mu_k| = 0.25$ and $|\mu_k| = 0.35$, the projection bands calculated using comsol are shown by circles in Fig. 8 (a) and (b), respectively. We can see that for both cases there is one surface state for one Bloch wave vector at the inner interface, see the red circles in Fig. 8. And the profiles of surface state dispersions are very different from those of traditional valley polarized surface sates which are predicted by the difference of the valley Chern numbers of two sides [47, 48].

To calculate the surface states using the effective medium theory, we firstly obtained the effective constitutive parameters according to Eq. (13) for $|\mu_k| \neq 0.3$. When the opened band gap is small, the effective constitutive parameters can be still written into the linear forms as Eq. (14) and neglect the higher order terms $o(\Delta\tilde{\omega})$. Then the bulk band dispersions can be calculated by Eq. (21) and the results are shown by black lines in Fig. 8. We assume that the fields of the surface state for a given $k_y$ is expressed as

$$\begin{pmatrix} Z_0\bar{H}_x \\ Z_0\bar{H}_y \\ \bar{E}_z \end{pmatrix} = \begin{cases} \vec{\psi}_1 e^{ik_y y} e^{ikx-\kappa x}, x \geq 0 \\ \vec{\psi}_2 e^{ik_y y} e^{ikx+\kappa x}, x \leq 0 \end{cases}, (25)$$

where $x = 0$ is the interface of the two effective media which corresponds to the inner interface of the supercell, $\kappa > 0$ are real numbers characterizing the decaying rate, $\vec{\psi}_1$ ($\vec{\psi}_2$) is the eigenstate of effective medium at $x > 0$ ($x < 0$) corresponding to $k_x = k + i\kappa$ ($k_x = k - i\kappa$), which is obtained according to Eq. (21). The surface state band dispersion is then obtained by matching the boundary conditions, i.e. $\bar{H}_y$ and $\bar{E}_z$ should be continuous at $x = 0$.

The surface bands obtained by the effective medium calculation are shown by the red lines in Fig. 8. The surface states at the boundary of two semi-infinite domains are calculated and they correspond to the inner interface surface states. We can see that the results agree well with those based on real periodic structure (red circles), especially when they are close to the $\Gamma$ point. So the effective parameters describe not only the bulk properties of the photonic crystal, but also faithfully predict the bands of surface states when the wave vector is small. For larger $k_y$, the decay length of the surface wave becomes shorter as $\kappa$ increases. As expected, the effective medium approach is less accurate for describing the surface states with larger $k_y$. This is because the effective medium approximation breaks down if the eigenfield is not uniform and the effective medium description is also not a good approximation for waves penetrating too shallow into the metamaterial.

## V. Non-Hermitian perturbations and magneto-optical complex conjugate metamaterials

Different from Hermitian perturbations which lift the degeneracy at the Dirac point, non-Hermitian perturbations will coalesce the linear dispersion bands and form EPs [25]. When the system becomes non-Hermitian, the effective constitutive parameters become complex. A special case is the CCM whose effective relative permittivity and permeability are complex conjugate to each other, i.e, $\varepsilon = \alpha e^{i\varphi}, \mu = \beta e^{-i\varphi}$, where $\alpha, \beta$ and $\varphi$ are non-zero real numbers [35-37]. The EM energy is balanced for a plane wave propagating inside the infinite CCM because the refractive index is pure real. However, lossy or gain mode can be excited when the EM wave is scattered by the boundaries since the impedance of CCM is complex. In a CCM slab with an appropriate width, either the CPA or lasing modes can be supported [35]. The scheme for realizing a CCM was firstly proposed by Bai et al., [36] who used a lattice of core-shell cylinders with the core and shell made of lossy and gain materials, respectively. Later, X. Cui et al., [26] demonstrated that CCMs for a wide frequency range can be realized by using a non-PT symmetric photonic crystal supporting a ring of EPs locating on the real frequency. Here we will extend the CCM to the magneto-optical one whose permeability is generalized from a scalar to a tensor possessing imaginary off-diagonal components.

### A. Formulation of the non-Hermitian Hamiltonian

Consider that the permittivity of the inclusions and background become complex, the non-Hermitian Hamiltonian can be formulated according to the wave equation for the electric field,

$$\nabla_\perp \times (\tilde{\vec{\mu}}^{-1}(\mathbf{r}) \cdot \nabla_\perp \times E_z(\mathbf{r})\hat{z}) + \frac{\omega^2}{c^2}\tilde{\varepsilon}(\mathbf{r})E_z(\mathbf{r})\hat{z} = 0, (26)$$

where $\tilde{\varepsilon}(\mathbf{r}) = \varepsilon(\mathbf{r}) + i\varepsilon_i(\mathbf{r})$ is the location dependent complex permittivity. Similarly, the Bloch state of the non-Hermitian system can be expanded in series of the normalized eigenstates $\{w_n(\mathbf{r})\}$ of the Hermitian system ($\varepsilon_i(\mathbf{r}) = 0$) as

$$\tilde{w}_{n\mathbf{k}}(\mathbf{r}) = \sum_m \beta_{nm} w_{m\mathbf{k}}(\mathbf{r}), (27)$$

where $w_{n\mathbf{k}}$ is the cell function of the $n$th band (shown in Fig.1 (b)), $\mathbf{k}$ denotes Bloch wave vector, and $\beta_{nm}$ is the expansion coefficient. Substituting Eq. (27) into Eq. (26), we have

$$-\sum_m \frac{\omega_m^2}{c^2}\varepsilon_r w_{m\mathbf{k}}(\mathbf{r}) + \frac{\omega^2}{c^2}\varepsilon_c \sum_n \beta_{nm} w_{m\mathbf{k}}(\mathbf{r}) = 0, (28)$$

where $\omega_m$ is the frequency of the $m$th band of the Hermittian system (shown in Fig. 1(b)). Multiplying Eq.(28) by $w_{n\mathbf{k}}^*(\mathbf{r})$ and integrating within a unit cell, we obtain the effective eigen-equation:

$$\tilde{H}_1^{-1} \cdot \tilde{H}_2 \cdot \tilde{\psi} = \tilde{H} \cdot \tilde{\psi} = \left(\frac{\omega}{c}\right)^2 \tilde{\psi}, (29)$$

where $\tilde{\psi} = \{\beta_{nm}\}^T$, $\tilde{H}$ is the non-Hermitian Hamiltonian, and components of $\tilde{H}_1, \tilde{H}_2$ are given by

$$\tilde{H}_{1,nm} = \iint w_{n\mathbf{k}}^*(\mathbf{r})\tilde{\varepsilon}(\mathbf{r})w_{m\mathbf{k}}(\mathbf{r})d\mathbf{r},$$
$$\tilde{H}_{2,nm} = \left(\frac{\omega_{m\mathbf{k}}}{c}\right)^2 \delta_{mn} \iint w_{m\mathbf{k}}^*(\mathbf{r})\varepsilon(\mathbf{r})w_{m\mathbf{k}}(\mathbf{r})d\mathbf{r}. \quad (30)$$

**B. Realization of magneto-optical complex conjugate metamaterial**

To realize the magneto-optical CCM, we consider that the permittivity of the inclusions and background in our photonic crystal are expressed as $\varepsilon_r + i\varsigma$ and $1 + il_r\varsigma$, respectively, where $l_r < 0$ is a real number denoting the loss-gain ratio. Only considering the contribution from the second and third bands using Eq. (25) to the effective Hamiltonian in Eq.(29), we obtain

$$\hat{H}_1 = \begin{pmatrix} 1 & 0 \\ 0 & 1 \end{pmatrix} + \varsigma \begin{pmatrix} \tau_2 & v \\ -v^* & \tau_3 \end{pmatrix}, \quad \hat{H}_2 = \frac{1}{c^2}\begin{pmatrix} \omega_2^2 & 0 \\ 0 & \omega_3^2 \end{pmatrix}, \quad (31)$$

where

$$\tau_i(\mathbf{k}) = \iint_{S_c} w_{i\mathbf{k}}^*(\mathbf{r})w_{i\mathbf{k}}(\mathbf{r})d\mathbf{r} + l_r \iint_{S_s} w_{i\mathbf{k}}^*(\mathbf{r})w_{i\mathbf{k}}(\mathbf{r})d\mathbf{r} = F_c^i(\mathbf{k}) + l_r F_b^i(\mathbf{k}),$$
$$v(\mathbf{k}) = \iint_{S_c} w_{2\mathbf{k}}^*(\mathbf{r})w_{3\mathbf{k}}(\mathbf{r})d\mathbf{r} + l_r \iint_{S_s} w_{2\mathbf{k}}^*(\mathbf{r})w_{3\mathbf{k}}(\mathbf{r})d\mathbf{r}, \quad (32)$$

with $S_c$ and $S_b$ denoting the regions inside and outside the elliptical cylinder within the unit cell. Then the eigenvalues for the effective Hamiltonian $\hat{H} = \hat{H}_1^{-1} \cdot \hat{H}_2$ are given by

$$\omega^2 = \frac{\Delta_1 \pm \sqrt{\Delta_1^2 - 4\omega_2^2\omega_3^2\Delta_2}}{2\Delta_2},$$
$$\Delta_1 = (\omega_2^2 + \omega_3^2 + \varsigma\omega_3^2\tau_2 + \varsigma\omega_2^2\tau_3), \quad (33)$$
$$\Delta_2 = 1 + \varsigma(\tau_2 + \tau_3) + \varsigma^2(\tau_2\tau_3 + vv^*),$$

and the EPs emerge at $\Delta_1^2 - 4\omega_2^2\omega_3^2\Delta_2 = 0$. If $l_r$ is well selected, in the following we will see $\Delta_1$, $\Delta_2$ are almost purely real and the EPs just separate the unbroken ($\Delta_1^2 - 4\omega_2^2\omega_3^2\Delta_2 > 0$) and broken ($\Delta_1^2 - 4\omega_2^2\omega_3^2\Delta_2 < 0$) phases.

The refractive index of the homogenized medium is defined as $kc/\omega$, which is real only when the frequencies are real, namely inside the unbroken phase. Note that the broken phase corresponds to the k-gap for a monochromatic waves, which can be seen from the complex Bloch k bands for the given real frequencies as shown in Fig. 10(a). Therefore, the broken phase will not be considered in the effective medium calculation. According to Eq. (33), a situation for fulfilling the condition that the eigenvalues are real when $\Delta_1^2 - 4\omega_2^2\omega_3^2\Delta_2 > 0$ is $\tau_2(\mathbf{k}) = \tau_3(\mathbf{k}) = 0$.

The parameter $\tau_i(\mathbf{k})$ in the effective Hamiltonian can be separated into two parts (see Eq. (32)). In Fig. 9(a), we plot the ratios of the two parts $F_c^2(\mathbf{k})/F_b^2(\mathbf{k})$ (black solid line) and $F_c^3(\mathbf{k})/F_b^3(\mathbf{k})$ (red dashed line) for different Bloch wave vector $\mathbf{k}$. We see that along ΓY direction, $F_c^2(\mathbf{k})/F_b^2(\mathbf{k}) \approx F_c^3(\mathbf{k})/F_b^3(\mathbf{k}) \approx 0.1449$ for a wide range of $k_y$. Therefore if we let $l_r = -0.1449$, the frequencies in the unbroken phase will be almost real for a wide range of $k_y$ since $\tau_2(\mathbf{k}) \approx \tau_3(\mathbf{k}) \approx 0$. For $\varsigma = 0.1$, we plot the real and imaginary parts of the second and third bands in Figs. 9(b) and 9(c). We can see that in the unbroken phase region where $|k_y|a \geq 0.01\pi$, the frequencies are almost real. The bands calculated using the 2-band Hamiltonian Eq. (31) (lines) agree well with the result of full wave simulations (circles), demonstrating the validity of Eq. (31). The slight discrepancy is due to the ignoring of other bands.

The effective constitutive parameters can be calculated using the boundary effective medium approach, see Eq. (13). As the effective medium here is applied to describe the behavior of waves with real frequencies, we should use the eigenstates with complex Bloch wave vectors but real-valued frequencies to retrieve the effective material parameters [26]. We refer to the results of the scenario of complex Bloch k versus the real frequency as complex Bloch k bands. The complex Bloch k bands as well as the corresponding eigenstates are calculated using the weak-form PDE module in COMSOL [50-52] (see details in the Appendix B), and the obtained band dispersion is shown in Fig. 10(a). We can see that there is a k-gap within the range $|k_y|a < 0.01\pi$ along $\Gamma Y$ direction, which just corresponds to the broken phase in Fig. 9 (b) and (c).

The real and imaginary parts of the obtained effective constitutive parameters are shown in Figs. 10(b) and 10(c). We can see that both the real and imaginary parts are almost linearly proportional to $\omega - \omega_D$. And $\mu_{ek} = \mu_{ek}'$ are still purely real numbers. However, different from traditional isotropic CCMs, the effective permittivity is no longer complex conjugate to any permeability tensor component. But the refractive index $n_y$ is real, as we can see from Fig. 11 (a). If we define an equivalent permeability for the wave propagating along y direction as

$$\tilde{\mu}_y = \frac{\mu_{ex}\mu_{ey} - \mu_k^2}{\mu_{ey}}, (34)$$

then $\varepsilon_{ez}$ and $\tilde{\mu}_y$ are complex conjugate to each other up to a real coefficient, namely they can be written as $\varepsilon_{ez} = |\varepsilon_{ez}|e^{i\varphi}, \tilde{\mu}_y = |\tilde{\mu}_y|e^{-i\varphi}$. This can be verified from Fig. 11(b), where a good agreement between the phases $\arg[\varepsilon_{ez}]$ and $-\arg[\tilde{\mu}_y]$ in a broad frequency range is seen.

## C. CPA and lasing

We consider two counter-propagating plane waves with the same frequency $\omega$ incident on a magneto-optical CCM slab with thickness $d$ embedded in air, as shown in Fig. 12(a). The incident plane waves are $E_z$ polarized and propagate along $y$ direction. Then the electric fields outside the magneto-optical CCM slab are expressed as

$$E_z = \begin{cases} a_1 e^{ik_0 y} + b_1 e^{-ik_0 y}, & y \leq -d/2 \\ b_2 e^{ik_0 y} + a_2 e^{-ik_0 y}, & y \geq -d/2 \end{cases}, (35)$$

where $k_0 = \omega/c$ is the wavenumber in air, $a_i, b_i, i=1,2$ are coefficients of the propagating waves. In terms of scattering matrix $\tilde{S}$, we can write [36]

$$\begin{pmatrix} b_2 \\ b_1 \end{pmatrix} = \tilde{S} \begin{pmatrix} a_1 \\ a_2 \end{pmatrix} = \frac{1}{M_{22}} \begin{pmatrix} 1 & M_{12} \\ M_{12} & 1 \end{pmatrix} \begin{pmatrix} a_1 \\ a_2 \end{pmatrix}, (36)$$

where

$$M_{12} = \frac{i}{2}(\frac{1}{Z_{ey}} - Z_{ey})\sin(n_y k_0 d),$$
$$M_{22} = e^{ik_0 d}[\cos(n_y k_0 d) - \frac{i}{2}(\frac{1}{Z_{ey}} + Z_{ey})\sin(n_y k_0 d)], \quad (37)$$

with $n_y$ and $Z_{ey}$ are the refractive index and impedance of the slab, respectively. The eigenvalues and eigenvectors of the scattering matrix are easily obtained as $\lambda_\pm = (1 \pm M_{12})/M_{22}$ and $\vec{\varphi}_\pm = (1, \pm 1)^T$. Therefore, the outgoing waves are expressed as

$$\begin{pmatrix} b_2 \\ b_1 \end{pmatrix} = \frac{1}{2}[(a_1 + a_2)\lambda_+ \vec{\varphi}_+ + (a_1 - a_2)\lambda_- \vec{\varphi}_-]. (38)$$

To achieve the CPA mode where the outgoing wave vanishes $b_1 = b_2 = 0$, either one of the following conditions are fulfilled: (1) $a_1 = -a_2$ (out-of-phase) and $\lambda_- = 0$, and (2) $a_1 = a_2$ (in-phase) and $\lambda_+ = 0$, which requires

$$\frac{i}{2}(\frac{1}{Z_{ey}} - Z_{ey})\sin(n_y k_0 d) = \pm 1. (39)$$

To achieve the lasing mode where the outgoing wave is divergent $|b_{1,2}| \gg |a_{1,2}|$, either $\lambda_+$ should be divergent for in-phase incidence or $\lambda_-$ should be divergent for

out-of-phase incidence, which requires

$$\cos(n_y k_0 d) = \frac{i}{2}(\frac{1}{Z_y} + Z_y)\sin(n_y k_0 d). \quad (40)$$

Because $n_y$ is almost real, Eqs. (39) and (40) are valid only when $Z_{ey}$ is pure imaginary, i.e., $\arg[\varepsilon_{ez}] = -\arg[\tilde{\mu}_y] = -\pi/2$, which corresponds to $\omega a/(2\pi c) \approx 0.55$, see Fig. 11(b).

At $\omega a/(2\pi c) \approx 0.55$, we plotted the absolute values of the eigenvalues $|\lambda_{\pm}|$ varying with the width of the metamaterial slab $d$ in Fig. 12(b). We can see that $|\lambda_-|$ is close to zero when $d = 30a$ and $|\lambda_+|$ reaches a maximum at $d = 65a$. Therefore, the CPA mode is achieved when the counter-propagating plane waves are out-of-phase ($a_1 = -a_2$) and the slab has 30 layers, while the lasing mode is achieved when the counter-propagating plane waves are in-phase ($a_1 = a_2$) and the slab has 65 layers. If we change the sign of $\varsigma$ so that the cylinder is gain and the background is lossy, the effective constitutive parameters become $\varepsilon_{ez}^*, \tilde{\mu}_e^*$. Then $|\lambda_-|$ will reach a peak at $d = 30a$ and $|\lambda_+|$ will form a deep at $d = 65a$, as we can see in Fig. 12(c). Thus the CPA mode is achieved when the plane waves are in-phase and the slab has 65 layers, while the lasing mode is achieved when the plane waves are out-of-phase and the slab has 30 layers.

In Fig. 13, we show the CPA effect induced by two counter-propagating plane waves of out-of-phase setting ($a_1 = -a_2 = E_0$) normally incident on the slab with a width $d = 30a$. The electric field distributions inside the effective medium slab and the slab composed of real structures are shown in Figs. 13(a) and 13(b), respectively. The

Poynting vectors shown by the white arrows in Fig. 13(a) are pointing inward, indicating that energies are absorbed by the slab. Inside the effective medium, the Poynting vectors are not parallel to y axis, see Fig. 13(a). This is because the x component of magnetic field $\bar{H}_x$ exists due to the off-diagonal term $\mu_{ek}$ in the permeability tensor. In Fig. 13(c), we show the electric field amplitudes $|E_z|$ along the dashed line in Fig. 13(b). We can see that $|E_z|$ outside the metamaterial are close to $|E_0|$, implying that the reflections of the boundaries are very small.

In Fig. 14, we show the lasing effect by considering two counter-propagating plane waves of out-of-phase setting ($a_1 = -a_2 = E_0$) normally incident on the slab with width $d = 30a$. The constitutive parameters of the slab are complex conjugate to those in Fig. 13. The electric field distributions for the effective medium slab and the slab composed of real structures are shown in Figs. 14(a) and 14(b), respectively. As we can see the Poynting vectors are now all pointing outward, indicating energies are delivered from the slab to the air background. Fig. 14(c) shows the field amplitudes along the dashed line in Fig. 14(b). It is clearly seen that the fields outside the slab are significantly amplified.

## VI. Conclusion

In summary, we realized an unpaired Dirac point at a relatively low frequency at the Brillouin zone center in a magneto-optical photonic crystal. We demonstrated that for frequencies near the unpaired Dirac cone, the system behaves as a homogenized magneto-optical zero-index medium for waves traveling in it, and such systems provide a new platform to study photonic pseudospin-1/2 systems. The effective permittivity as well as the determinant of the effective permeability tensor are simultaneously zero which can be regarded as a generalization of traditional double zero-index metamaterials. The Klein tunneling effect and Zitterbewegung effect in the homogenized effective medium were investigated and can be understood from the impedance of the medium and polarizations of the EM waves inside the medium, respectively. When the Dirac point is slightly lifted, the effective medium approach can also successfully predict the band dispersion of surface states near the Γ point confined on the interface between two such magnetic photonic crystals with opposite magnetization. Moreover, we formulated the non-Hermitian Hamiltonian of the photonic crystal when the permittivity of the inclusions and background of the photonic crystal became complex. We realized the EPs at the real frequency and showed that the non-Hermitian effective medium behaves like a magneto-optical CCM. We further discuss how to realize CPA and lasing modes using such magneto-optical CCM.

**Acknowledgments:** This work was supported by National Natural Science Foundation of China (NSFC) through Nos. 11904237, 11734012 and 11574218. Work done in Hong Kong is supported by Hong Kong RGC through grants 16303119 and N_HKUST608/17.

.

**Appendix A: Hamiltonian in Pauli matrices for the special case** $\gamma_x = \gamma_y, \mu_{x0} = \mu_{y0}$

For the sake of mathematical simplicity, we consider the special case that $\gamma_x = \gamma_y = \gamma_r, \mu_{x0} = \mu_{y0} = \mu_{r0}$ and set $\tilde{\omega}_D = 1$. Consider that $\mu_{k0} = \mu_{r0} + \delta$, then Eq. (21) can be written as

$$\vec{\vec{B}}^{-1} \cdot \vec{\vec{A}} \vec{\psi} = \Delta \tilde{\omega} \vec{\psi},$$

$$A = \begin{pmatrix} -\mu_{r0} & -i\mu_{k0} & \tilde{k}_y \\ i\mu_{k0} & -\mu_{r0} & -\tilde{k}_x \\ \tilde{k}_y & -\tilde{k}_x & 0 \end{pmatrix}, B = -\begin{pmatrix} \mu_{r0} + \gamma_r & i(\mu_{k0} + \gamma_k) & 0 \\ -i(\mu_{k0} + \gamma_k) & -\mu_{r0} & 0 \\ 0 & 0 & \gamma_z \end{pmatrix}. \quad (A1)$$

Imposing a unitary transform on $\vec{B}$ so that $\vec{B}'$ is diagonal

$$\vec{B}' = \vec{U} \cdot \vec{B} \cdot \vec{U}^\dagger = \begin{pmatrix} \gamma_e & & \\ & \gamma_r + \gamma_k + 2\mu_{r0} + \delta & \\ & & \gamma_r - \gamma_k - \delta \end{pmatrix}, \vec{U} = \begin{pmatrix} 0 & 0 & 1 \\ -i/\sqrt{2} & 1/\sqrt{2} & 0 \\ i/\sqrt{2} & 1/\sqrt{2} & 0 \end{pmatrix}. \quad (A2)$$

Then we further have

$$(\vec{B}')^{-1/2} \cdot \vec{U} \cdot \vec{A} \cdot \vec{U}^\dagger \cdot (\vec{B}')^{-1/2} \cdot (\vec{B}')^{1/2} \cdot \vec{U} \cdot \vec{\psi} = \Delta\tilde{\omega}(\vec{B}')^{-1/2} \cdot \vec{U} \cdot \vec{B} \cdot \vec{U}^\dagger \cdot (\vec{B}')^{-1/2} \cdot (\vec{B}')^{1/2} \cdot \vec{U} \cdot \vec{\psi}, \quad (A3)$$

which is rewritten as

$$\vec{A}' \cdot \vec{\psi}' = \Delta\omega\vec{\psi}', \quad (A4)$$

where

$$\vec{\psi}' = (\vec{B}')^{1/2} \cdot \vec{U} \cdot \vec{\psi}, \quad (A5)$$

and

$$\vec{A}' = (\vec{B}')^{-1/2} \cdot \vec{U} \cdot \vec{A} \cdot \vec{U}^\dagger \cdot (\vec{B}')^{-1/2}. \quad (A6)$$

Using the two eigenvectors at $\vec{k} = 0$ to formulate the Hamiltonian at $\vec{k} \neq 0$, viz $H_{\alpha\beta} = (\vec{\psi}_\alpha')^* \cdot \vec{A}' \cdot \vec{\psi}_\beta'$, we have

$$\tilde{H} = \begin{pmatrix} 0 & \dfrac{-\tilde{k}_x - i\tilde{k}_y}{\sqrt{2\gamma_e(\gamma_r - \gamma_k - \delta)}} \\ \dfrac{-\tilde{k}_x + i\tilde{k}_y}{\sqrt{2\gamma_e(\gamma_r - \gamma_k - \delta)}} & \dfrac{\delta}{\gamma_r - \gamma_k - \delta} \end{pmatrix} \quad (A7)$$

$$= -\dfrac{\tilde{k}_x}{\sqrt{2\gamma_e(\gamma_r - \gamma_k - \delta)}}\vec{\sigma}_x + \dfrac{\tilde{k}_y}{\sqrt{2\gamma_e(\gamma_r - \gamma_k - \delta)}}\vec{\sigma}_y + \dfrac{\delta}{2(\gamma_r - \gamma_k - \delta)}(\vec{\sigma}_0 - \vec{\sigma}_z),$$

where $\vec{\sigma}_0$ is the identity matrix and $\vec{\sigma}_{x,y,z}$ are Pauli matrices. According to Eq. (A7), we easily find the band gap size is $\delta/(\gamma_r - \gamma_k - \delta) \approx \delta/(\gamma_r - \gamma_k)$.

**Appendix B: Calculating the complex Bloch k bands using weak-form PDE**

**module in Comsol**

The effective constitutive parameters describe the optical properties of metamaterials at a given real frequency. So if the system is non-Hermitian, the eigenstates used for effective medium calculation should correspond to the real frequency and complex Bloch wave vectors. Here, we use the weak-form PDE module in comsol to calculate the complex Bloch k bands [48-51]. The weak-form expression used for the calculation is obtained as following.

According to the Maxwell's equation, the wave equation for out-of-plane electric field inside the magneto-optical medium is given by

$$\nabla_\perp \times (\bar{\bar{\mu}}^{-1}(\mathbf{r}) \cdot \nabla_\perp \times \mathbf{E}) - \varepsilon(\mathbf{r}) \frac{\omega^2}{c^2} \mathbf{E} = 0. \text{ (B1)}$$

According to Bloch theorem, the out-of-plane electric field can be expressed as

$$\mathbf{E} = w(\mathbf{r})e^{i\mathbf{k}\cdot\mathbf{r}}. \text{ (B2)}$$

Substituting Eq. (B2) into Eq. (B1), we arrive at

$$\nabla_\perp \times (\frac{\mu_r}{\mu_r^2 - \mu_k^2}\partial_y E_z + \frac{i\mu_k}{\mu_r^2 - \mu_k^2}\partial_x E_z, \frac{i\mu_k}{\mu_r^2 - \mu_k^2}\partial_y E_z - \frac{\mu_r}{\mu_r^2 - \mu_k^2}\partial_x E_z) - k_0^2 \varepsilon E_z = 0, \text{ (B3)}$$

which becomes

$$\frac{(ik_x, ik_y)}{\mu_r^2 - \mu_k^2} \times [\mu_r(w_y + ik_y w) + i\mu_k(w_x + ik_x w), i\mu_k(w_y + ik_y w) - \mu_r(w_x + ik_x w)]$$
$$+ \frac{\nabla}{\mu_r^2 - \mu_k^2} \times [\mu_r(w_y + ik_y w) + i\mu_k(w_x + ik_x w), i\mu_k(w_y + ik_y w) - \mu_r(w_x + ik_x w)] - k_0^2 \varepsilon w = 0, \quad \text{(B4)}$$

where $w_x = \partial w/\partial x, w_y = \partial w/\partial y$. Multiply the test function $\tilde{w}(\mathbf{r})$ and integrate within a unit cell, the weak-form is obtained as [49, 50]

$$weak(w) = \frac{-ik_y}{\mu_r^2 - \mu_k^2}[\mu_r(w_y + ik_yw) + i\mu_k(w_x + ik_xw)]\tilde{w} + \frac{ik_x}{\mu_r^2 - \mu_k^2}[i\mu_k(w_y + ik_yw) - \mu_r(w_x + ik_xw)]\tilde{w}$$

$$+ \frac{\tilde{w}_y}{\mu_r^2 - \mu_k^2}[\mu_r(w_y + ik_yw) + i\mu_k(w_x + ik_xw)] - \frac{\tilde{w}_x}{\mu_r^2 - \mu_k^2}[i\mu_k(w_y + ik_yw) - \mu_r(w_x + ik_xw)] - k_0^2 \varepsilon w\tilde{w}.$$

(B5)

In the simulations, the fields should be approximated with Lagrange interpolation elements. According to the Maxwell's equation, the magnetic fields are calculated by

$$\mathbf{H} = \frac{1}{i\omega\mu_0} \ddot{\mu}^{-1} \cdot [i\mathbf{k} \times (w\hat{z}) + \nabla_\perp \times (w\hat{z})]e^{i\mathbf{k}\cdot\mathbf{r}}. \quad (B6)$$

Concretely,

$$H_x = \frac{1}{i\omega\mu_0} \frac{\mu_k k_x w + i\mu_r k_y w - i\mu_k w_x + \mu_r w_y}{\mu_r^2 - \mu_k^2} e^{i\mathbf{k}\cdot\mathbf{r}},$$
$$H_y = \frac{1}{i\omega\mu_0} \frac{\mu_k k_y w - i\mu_r k_x w - i\mu_k w_y - \mu_r w_x}{\mu_r^2 - \mu_k^2} e^{i\mathbf{k}\cdot\mathbf{r}}.$$
(B7)

**References:**

1. M. I. Katsnelson, K. S. Novoselov, and A. K. Geim, Nat. Phys. 2, 620 (2006).

2. A. H. Castro Neto, F. Guinea, N. M. R. Peres, K. S. Novoselov, and A. K. Geim, Rev. Mod. Phys. 81, 109 (2009).

3. N. Stander, B. Huard, and D. Goldhaber-Gordon, Phys. Rev. Lett. 102, 026807 (2009).

4. M. I. Katsnelson, Eur. Phys. J. B 51, 157 (2006).

5. J. Cserti and G. David, Phys. Rev. B 74, 172305 (2006).

6. F. Dreisow, M. Heinrich, R. Keil, A. Tunnermann, S. Nolte, S. Longhi, and A. Szameit, Phys. Rev. Lett. 105, 143902 (2010).

7. X. Zhang, Phys. Rev. Lett. 100, 113903 (2008).

8. J. H. Bardarson, J. Tworzydlo, P. W. Brouwer, and C. W. J. Beenakker, Phys. Rev. Lett. 99, 106801 (2007).

9. K. Nomura, M.Koshino, and S. Ryu, Phys. Rev. Lett. 99, 146806 (2007).

10. Y. B. Zhang, Y. W. Tan, H. L. Stormer, and P. Kim, Nature (London) 438, 201 (2005).

11. K. S. Novoselov, Z. Jiang, Y. Zhang, S. V. Morozov, H. L. Stormer, U. Zeitler, J. C. Maan, G. S. Boebinger, P. Kim, and A. K. Geim, Science 315, 1379 (2007).

12. F. D. M. Haldane and S. Raghu, Phys. Rev. Lett. 100, 013904 (2008).

13. Z. Wang, Y. Chong, J. D. Joannopoulos, and M. Soljacic, Nature 461, 772 (2009).

14. Y. Plotnik, M. C. Rechtsman, D. Song, M. Heinrich, J. M. Zeuner, S. Nolte, Y. Lumer, N. Malkova, J. Xu, A. Szameit, Z. Chen, M. Segev, Nat. Mater. 13, 57 (2014).

15. D. Torrent and J. Sanchez-Dehesa, Phys. Rev. Lett. 108, 174301 (2012).

16. S.-Y. Yu, X.-C. Sun, X. Ni, Q. Wang, X.-J. Yan, C. He, X.-P. Liu, L. Feng, M.-H. Lu and Y.-F. Chen, Nat. Mater. 15, 1243 (2016).

17. X. Huang, Y. Lai, Z. H. Hang, H. Zheng and C. T. Chan, Nat. Mater. 10, 582 (2011).

18. F. Liu, X. Huang, and C. T. Chan, Appl. Phys. Lett. 100, 071911 (2012).

19. M. Dubois, C. Shi, X. Zhu, Y. Wang and X. Zhang, Nat. Commun. 8, 14871 (2017) .

20. R. Shen, L. B. Shao, B.Wang, and D. Y. Xing, Phys. Rev. B 81, 041410 (2010).

21. D. F. Urban, D. Bercioux, M. Wimmer, and W. Hausler, Phys. Rev. B 84, 115136 (2011).

22. A. Fang, Z. Q. Zhang, S. G. Louie, and C. T. Chan, Phys. Rev. B 93, 035422

(2016).

23. A. Fang, Z. Q. Zhang, S. G. Louie, and C. T. Chan, Proc. Natl. Acad. Sci. USA 114, 4087 (2017).

24. J. Mei, Y. Wu, C. T. Chan, and Z.-Q. Zhang, Phys. Rev. B 86, 035141 (2012).

25. B. Zhen, C. W. Hsu, Y. Igarashi, L. Lu, I. Kaminer, A. Pick, S.-L. Chua, J. D. Joannopoulos and Marin Soljacic, Nature 525, 354 (2015).

26. X. Cui, K. Ding, J. Dong, and C. T. Chan, Nanophotonics 9, 195–203 (2020).

27. L. Lu, C. Fang, L. Fu, S. G. Johnson, J. D. Joannopoulos, and M. Soljacic, Nat. Phys. 12, 337 (2016).

28. H. B. Nielsen and M. Ninomiya, Nucl. Phys. B 193, 173 (1981).

29. K. Sakoda, Opt. Express 20, 9925 (2012)

30. Y. Li, Y. Wu, and J. Mei, Appl. Phys. Lett. 105, 014107 (2014).

31. Z.-G. Chen, X. Ni, Y. Wu, C. He, X.-C. Sun, L.-Y. Zheng, M.-H. Lu and Y.-F. Chen, Sci. Rep. 4, 4613 (2014).

32. X. Ni, D. Purtseladze, D. A. Smirnova, A. Slobozhanyuk, A. Alu, A. B. Khanikaev, Sci. Adv. 4, eaap8802 (2018).

33. X. Zhou, D. Leykam, U. Chattopadhyay, A. B. Khanikaev, and Y. D. Chong, Phys. Rev. B 98, 205115 (2018).

34. G.-G. Liu, P. Zhou, Y. Yang, H. Xue, X. Ren, X. Lin, H.-X. Sun, L. Bi, Y. Chong, and B. Zhang, Nat. Commun. 11, 1873 (2020).

35. D. Dragoman, Opt. Commun. 284, 2095 (2011).

36. P. Bai, K. Ding, G. Wang, J. Luo, Z.-Q. Zhang, C. T. Chan, Y. Wu, and Y. Lai, Phys. Rev. A 94, 063841 (2016).

37. Y. Xu, Y. Fu, H. Chen, Opt. Express 25, 4952 (2017).

38. Y. Wu, Opt. Express 22, 1906 (2014).

39. D. R. Smith and J. B. Pendry, J. Opt. Soc. Am. B 23, 391 (2006).

40. A. Alu, Phys. Rev. B 84, 075153 (2011).

41. www.comsol.com.

42. O. Klein, Zeitschrift fr Physik 53, 157 (1929).

43. S. Longhi, Phys. Rev. B 81, 075102 (2010).

44. T. Ozawa, A. Amo, J. Bloch, I. Carusotto, Phys. Rev. A 96, 013813 (2017).

45. J.D. Jackson, Classical Electrodynamics (1999).

46. A. B. Khanikaev, S. H Mousavi, W.-K. Tse, M. Kargarian, A. H. MacDonald and G. Shvets, Nat. Mater. 12, 233 (2013).

47. T. Ma and G. Shvets, New J. Phys. 18, 025012 (2016).

48. J.-W. Dong, X.-D. Chen, H. Zhu, Y. Wang and X. Zhang, Nat. Mater. 16, 298 (2017).

49. T. Ma. and G. Shvets, Phys. Rev. B 95, 165102 (2017).

50. M. Davanc, Y. Urzhumov, and G. Shvets, Opt. Express 15, 9681 (2007).

51. M. Davanc, Y. Urzhumov, and G. Shvets, Opt. Express 19, 19027 (2011).

52. N. Wang, S. Wang, Z.-Q. Zhang, and C. T. Chan, Phys. Rev. B 98, 045426 (2018).

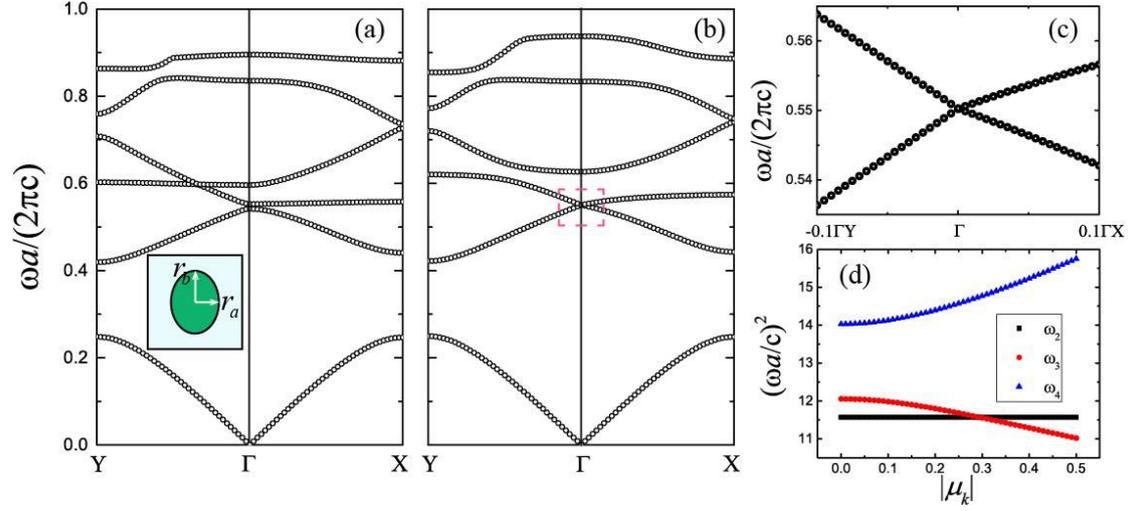

**Fig.1.** Band structures of the TRS preserved ($\mu_k = 0$) (a) and TRS broken ($\mu_k = 0.3$) (b) photonic crystals. The inset in (a) shows the unit cell. The semi-minor and semi-major axes of the cylinder are $r_a = 0.17a, r_b = 1.205 r_a$, the relative permiitivity of the cylinder is $\varepsilon_r = 12.5$ and the diagonal terms of the permeability tensor is $\mu_r = 1$. (c) The zoomed in view of a piece of the second and third bands in (b) near Γ point (marked by the dashed rectangle). (d) Evolution of the second (black), third (red) and fourth (blue) bands at Γ point as $\mu_k$ changes. Using the eigenstates of the TRS preserved system we obtain that $p_{23} = -p_{32} = p_{24} = -p_{42} \approx 0, p_{34} = -p_{43} = 0.32$ and used them for calculating the data points in (d).

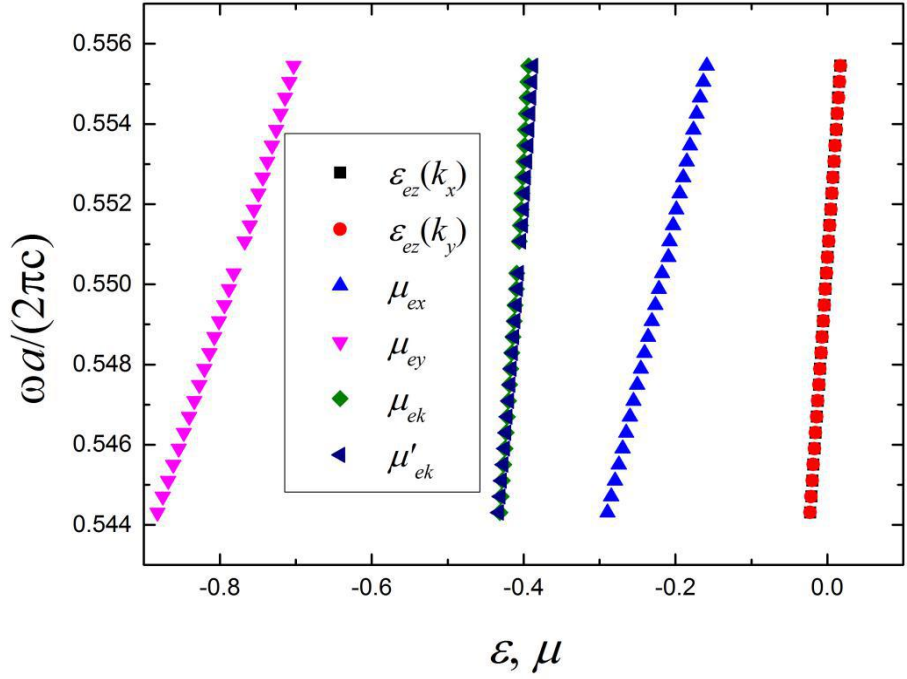

**Fig. 2.** Effective constitutive parameters calculated using the boundary effective medium approach. $\varepsilon_{ez}(k_x)$ and $\varepsilon_{ez}(k_y)$ denote the effective permittivity are calculated by the first and second sub-equations in Eq. (13) which are using the eigenstates whose Bloch wave vector is along ΓX and ΓY directions, respectively. By linear fitting the symbol lines, we obtain that $\varepsilon_{ez} \approx 0.565\Delta\tilde{\omega}$, $\mu_{ex} \approx -0.213 + 1.868\Delta\tilde{\omega}$, $\mu_{ey} \approx -0.776 + 2.571\Delta\tilde{\omega}$ and $\mu_{ek} = \mu_{ek}' \approx -0.407 + 0.591\Delta\tilde{\omega}$.

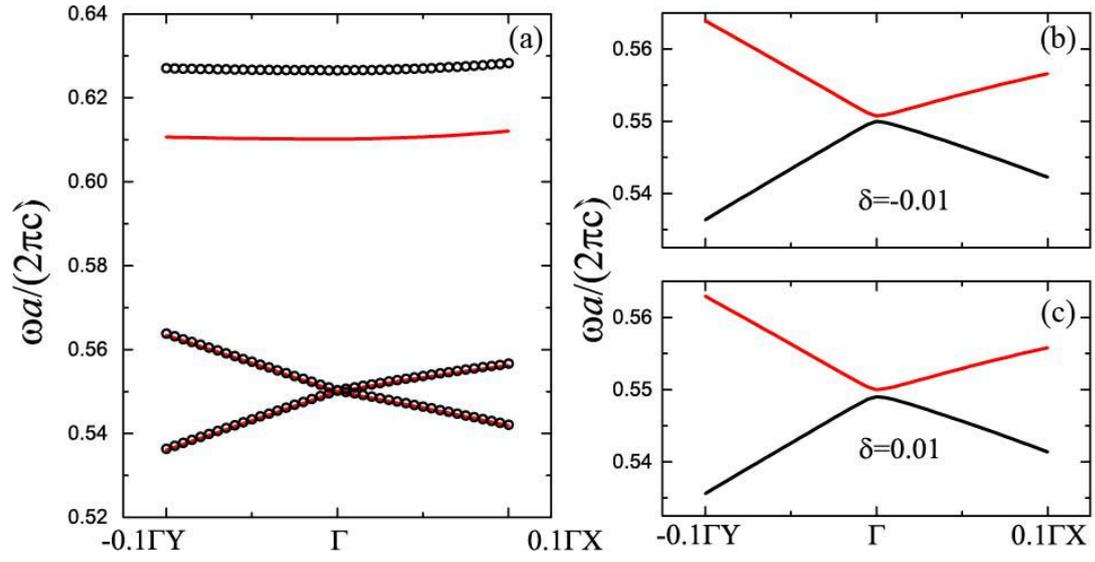

Fig. 3. (a) The second, third and fourth bands of the photonic crystal (circles) which are calculated using comsol and the bands of the effective medium (lines) which are calculated using Eq. (21). (b) and (c) the Dirac point is lifted when $\mu_{k0} = \pm(\sqrt{\mu_{x0}\mu_{y0}} + \delta), \delta \neq 0$.

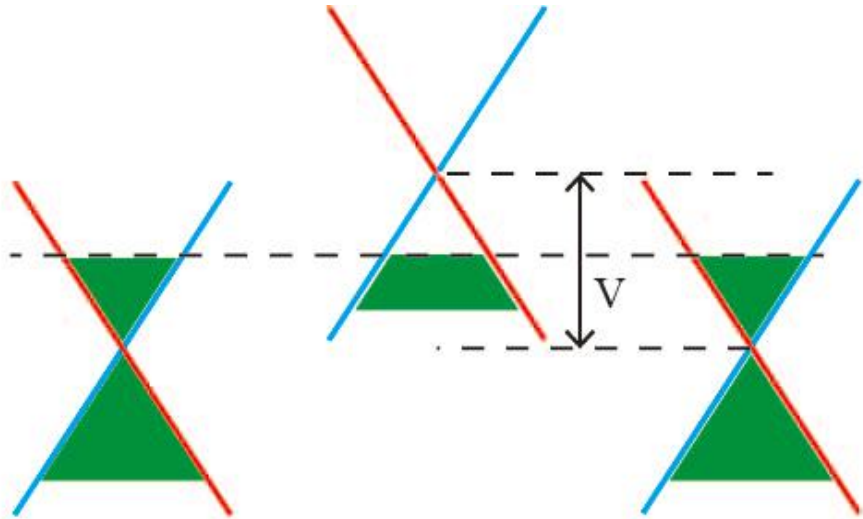

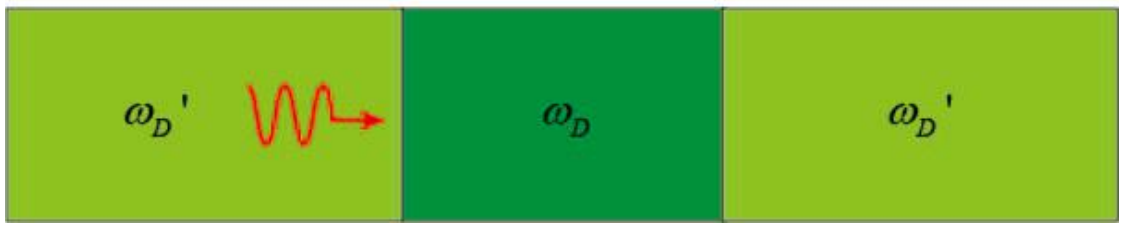

Fig.4. The schematic diagram of EM plane wave penetrating a square potential barrier. The effective medium slab with the Dirac point frequency $\omega_D{'}$ embedded in an effective medium background with the Dirac point frequency $\omega_D$ forms a square potential barrier $V = \omega_D - \omega_D{'}$.

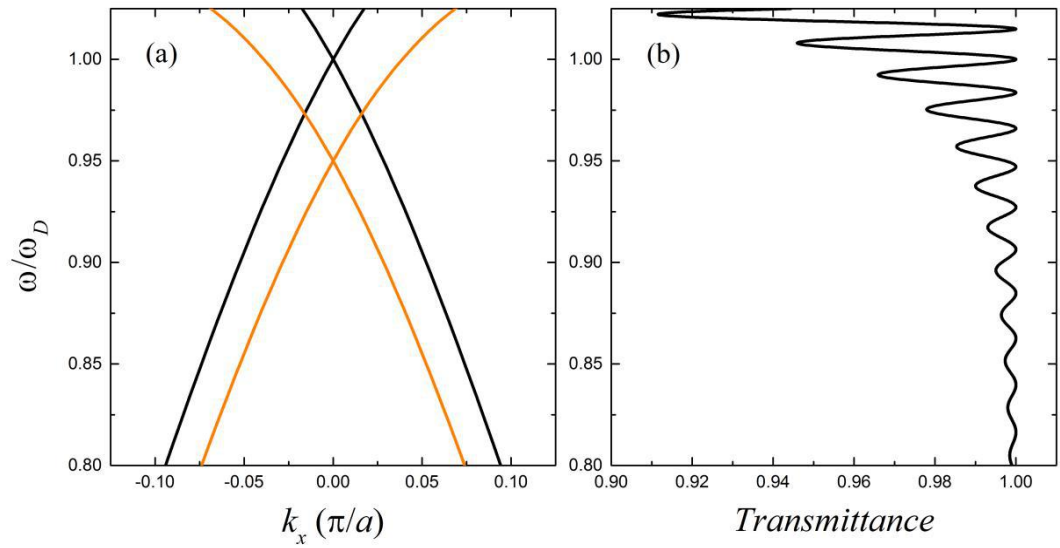

Fig. 5. (a) Dispersion relations of the effective medium slab (black) and the effective medium background (orange). (b) Transmittance of the EM plane waves at different frequencies. The width of the slab is set as $100a$.

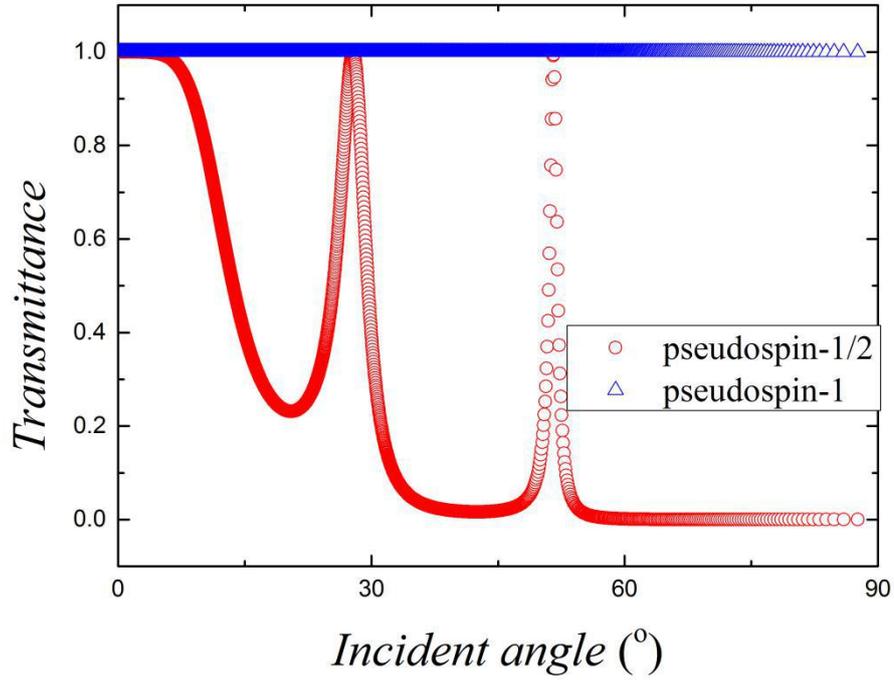

Fig. 6. Transmittance of the EM plane waves with the same frequency $\omega = (\omega_D + \omega_D')/2$ for different incident angles. The red circles show the result for the pseudospin-1/2 effective medium case, while the blue triangles show the result for the pseudospin-1 ( $\mu_{x0} = \mu_{y0} = \mu_{k0} = \gamma_k = 0$ ) effective medium case. The width of the slab is set as $200a$.

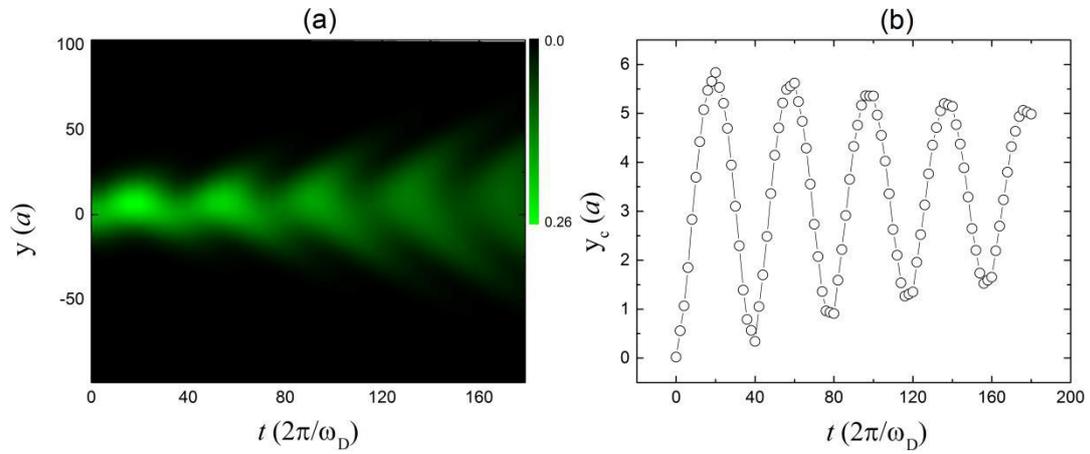

Fig. 7. Temporal Zitterbewegung effect. (a) The effective energy distribution along y direction when time grows. (b) The mass center $y_c$ as function of time $t$. The wavepacket has $k_x = 0.3/a$ and $\delta_y = 0.1/a$.

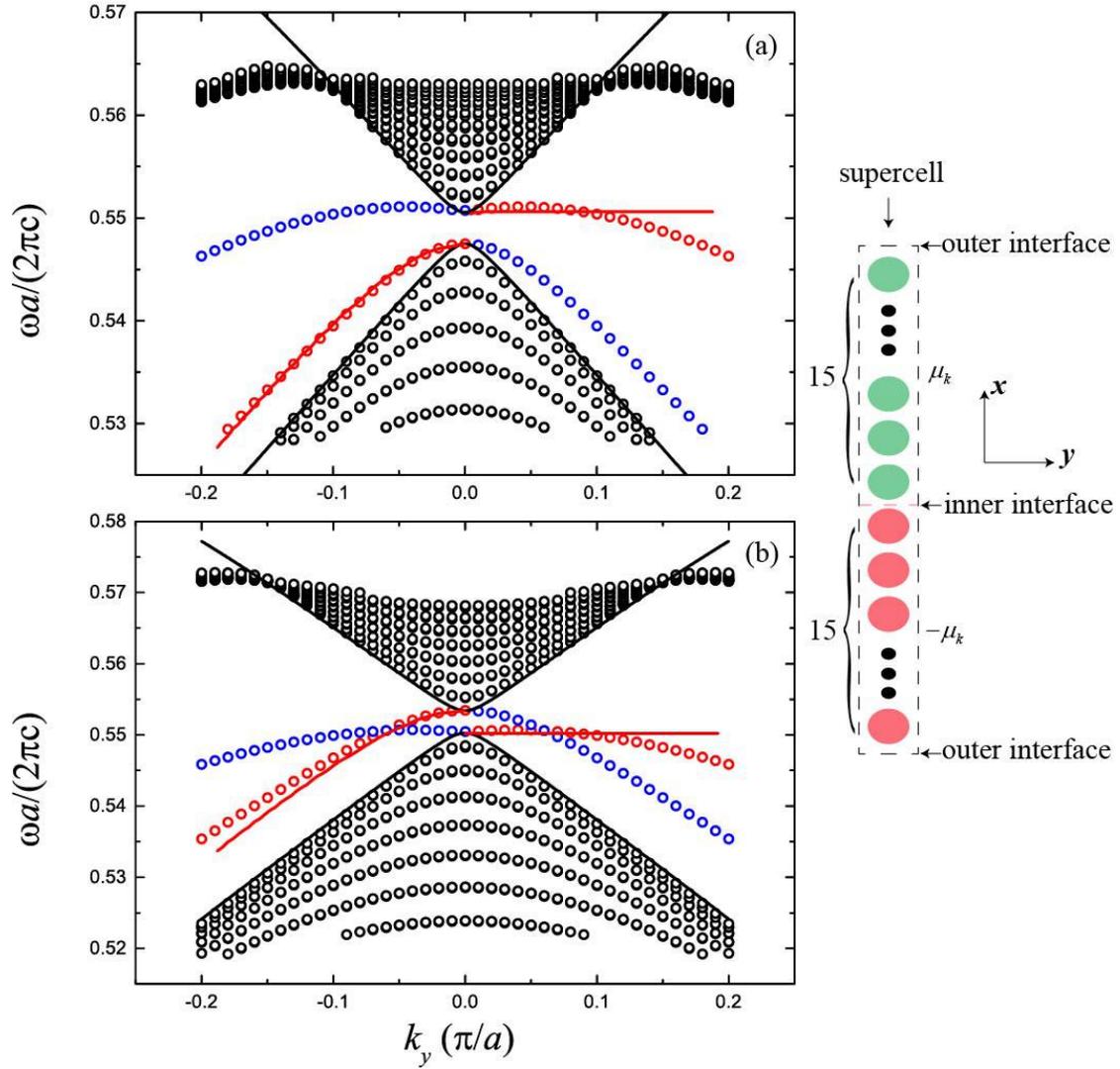

Fig. 8. The projection bands by supercell calculation (circles) and the bulk and surface state dispersion of the effective medium (lines). The black circles denote the bulk bands, while the red and blue circles represent the surface bands at the inner and outer interfaces of the supercell, respectively. The supercell contains 30 layers in total, with the upper half and lower half cylinders magnetized oppositely. The bulk dispersion of the effective medium is calculated using Eq. (21) and shown by the black lines. For the effective medium, the surface state is calculated by Eq. (16) and matching boundary conditions, only the surface states at the inner interface are shown by the red lines. (a) $\mu_k = 0.25$ and (b) $\mu_k = 0.35$.

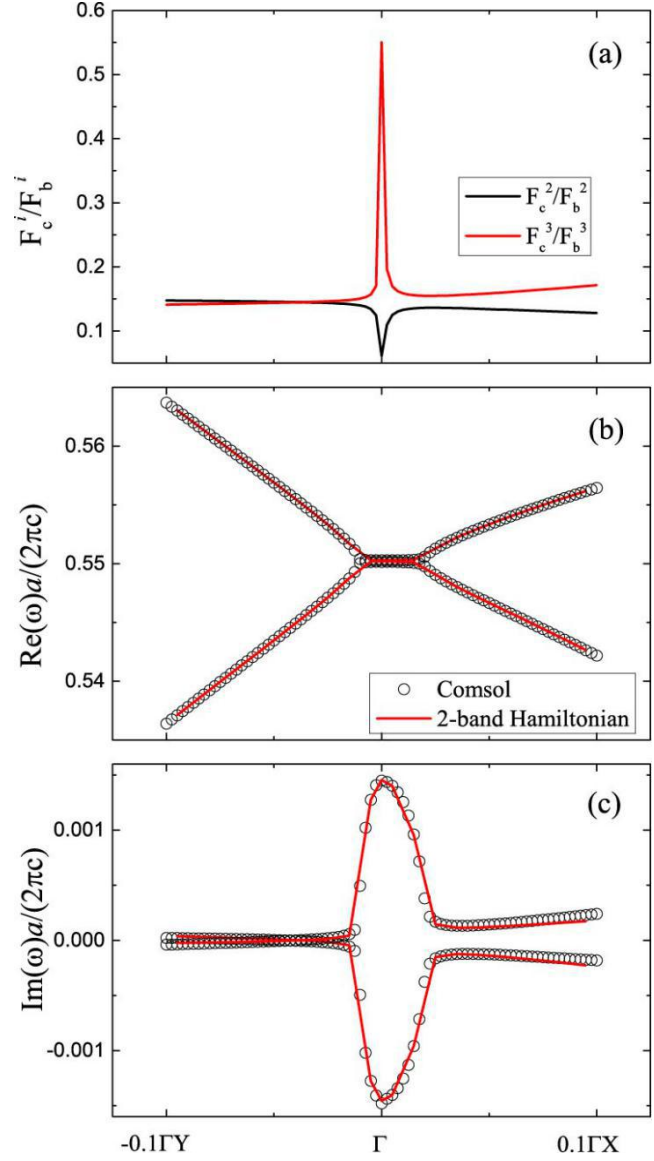

Fig. 9. (a) The integration ratio $F_c^2/F_b^2$ and $F_c^3/F_b^3$ for the second and third bands of the Hermitian photonic crystal as functions of the Bloch wave vectors. The real (b) and imaginary (c) parts of the second and third bands for the photonic crystal with non-Hermitian perturbations, which are calculated by COMSOL (circles) and 2-band Hamiltonian Eq. (28) (lines). The permittivity of the cylinder and background of the non-Hermitian system are $12.5+i\varsigma$ and $1+il_r\varsigma$, where $\varsigma=0.1, l_r=-0.14495$.

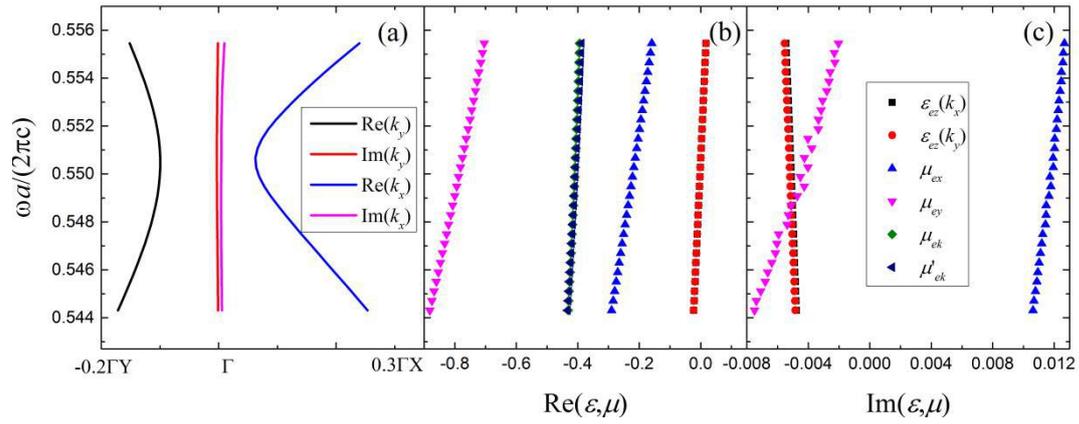

Fig. 10. (a) The complex Bloch wave vectors as functions of real frequencies. (b) The real parts of the effective constitutive parameters. (c) The imaginary parts of the effective constitutive parameters.

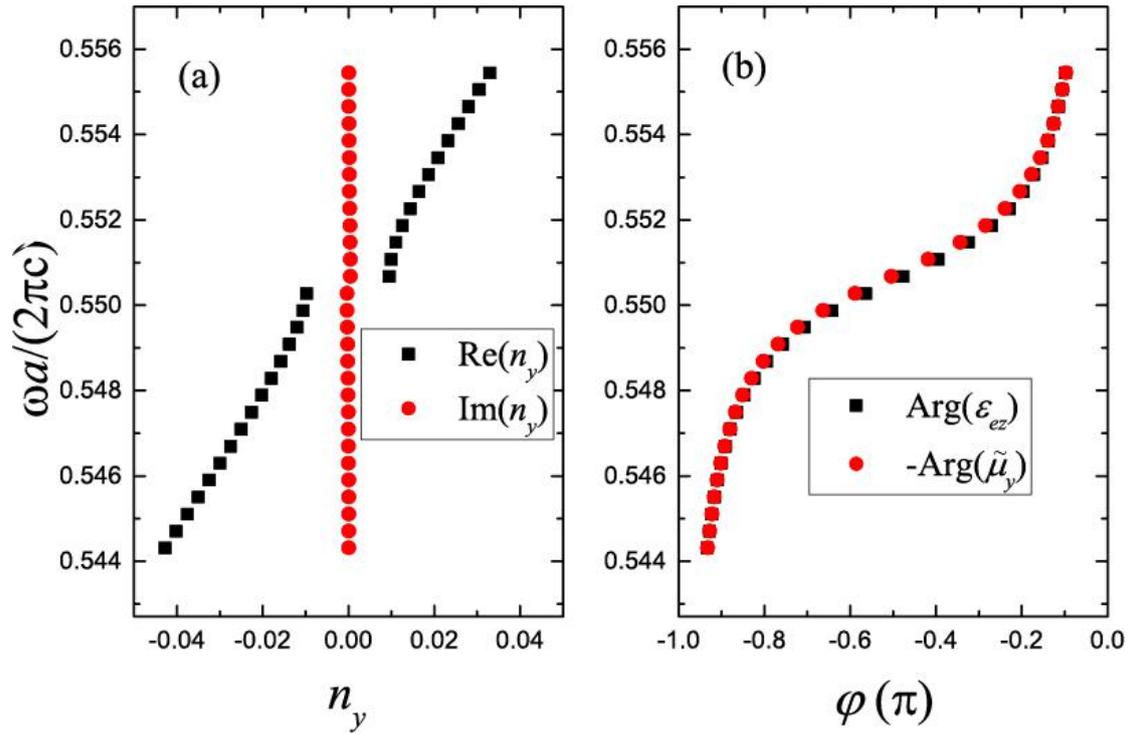

Fig.11. (a) Real and imaginary parts of the effective refractive index $n_y$ as functions of the frequencies. (b) The complex angles $\varphi$ of the effective permittivity $\varepsilon_{ez}$ and the equivalent permeability $\tilde{\mu}_y$ for different frequencies.

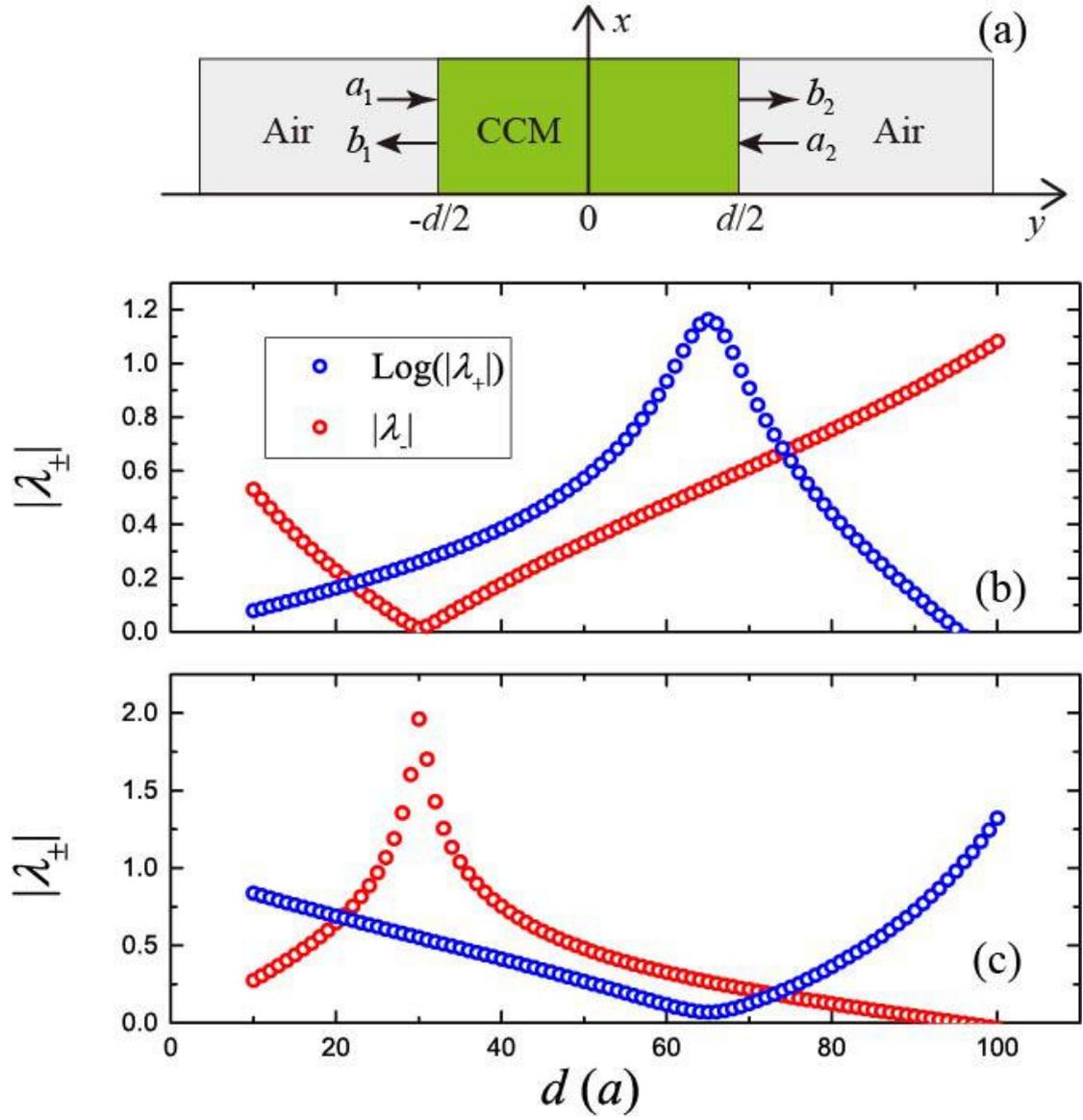

Fig. 12. (a) Schematic graph of two counter-propagating plane waves incident on a CMM slab. (b) Absolute values of the eigenvalues $\lambda_\pm$ as functions of the width of the metamaterial slab. (c) The gain and loss inside the materials are interchanged, i.e., $\varsigma = -0.1$. The frequency of the EM wave is $\omega a/(2\pi c) \approx 0.55$.

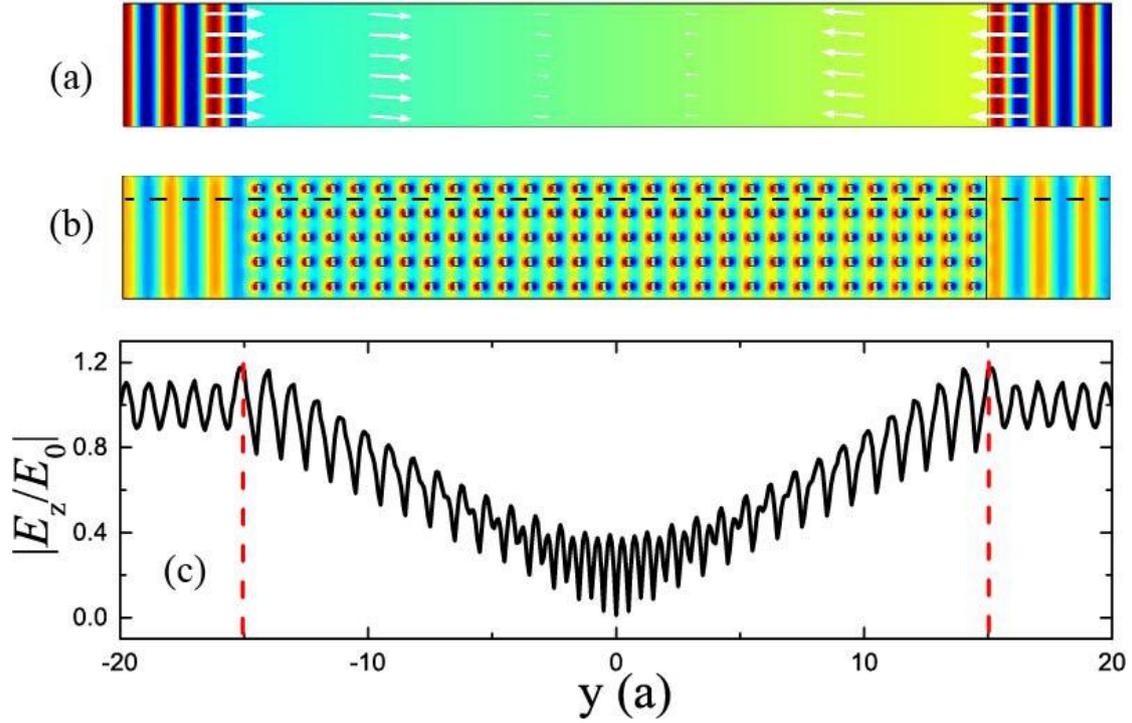

Fig. 13. (a) Real part of the electric field (color) and Poynting vectors (white arrows) for two counter-propagating plane waves normally incident on a magneto-optical CCM slab with a width $d=30a$. (b) Real part of Electric field for two counter-propagating plane waves normally incident on a slab with real structures. (c) The electric field amplitudes along the dashed line shown in (b). The red dashed lines denote the boundaries of the metamaterial. The two plane waves with frequency $\omega a/(2\pi a)=0.55$ are out-of-phase $a_1=-a_2=E_0$. For the metamaterial, the permittivity of the cylinder and background are $12.5+i\varsigma$ and $1-il_r\varsigma$, where $\varsigma=0.1, l_r=-0.1449$.

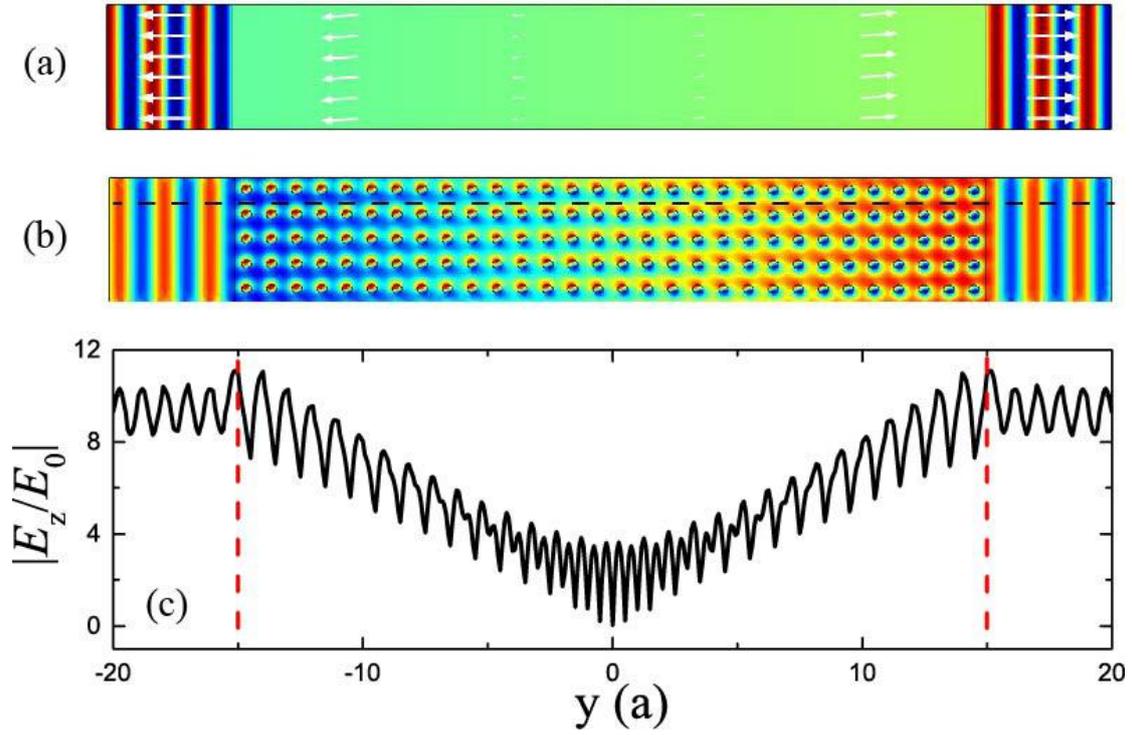

Fig. 14. Real part of electric field (color) and Poynting vectors (white arrows) for two counter-propagating plane waves normally incident on a magneto-optical CCM slab with a width $d=30a$. (b) Real part of electric field for two counter-propagating plane waves normally incident on a slab with real structures. (c) The electric field amplitudes along the dashed line shown in (b). The red dashed lines denote the boundaries of the metamaterial. The two plane waves with frequency $\omega a/(2\pi a)=0.55$ are out-of-phase $a_1=-a_2=E_0$. For the metamaterial, the permittivity of the cylinder and background are $12.5+i\varsigma$ and $1-il_r\varsigma$, where $\varsigma=-0.1, l_r=-0.1449$.